\documentclass[structabstract]{aa}

\usepackage{graphicx}
\usepackage[intlimits]{amsmath}
\usepackage{txfonts}
\usepackage{natbib}
\usepackage{color}
\bibpunct{(}{)}{;}{a}{}{,} 

\newcommand{\doverd}[2]{\frac{\partial #1}{\partial #2}}

\begin{document}
\title{Low-mass planets in nearly inviscid disks: \\
  Numerical treatment}

\author{
Wilhelm Kley\inst{\ref{inst1}}
\and Tobias~W.~A. M\"uller\inst{\ref{inst1}}
\and Stefan M. Kolb\inst{\ref{inst1}}
\and Pablo Ben\'\i tez-Llambay \inst{\ref{inst2}}
\and Fr\'ed\'eric Masset \inst{\ref{inst3}}
}

\institute{
     Institut f\"ur Astronomie \& Astrophysik, 
     Universit\"at T\"ubingen,
     Auf der Morgenstelle 10,
     72076 T\"ubingen,
     Germany
     \label{inst1}
\and
   Instituto de Astronom\'\i a Te\'orica y Experimental, IATE (CONICET),
   Observatorio Astron\'omico, Universidad Nacional de C\'ordoba, 
   Laprida 854, X5000BGR, C\'ordoba, Argentina
     \label{inst2}
\and
	Instituto de Ciencias F\'\i sicas,
        Universidad Nacional Aut\'onoma de M\'exico,
	Apdo. Postal 48-3,62251-Cuernavaca, Morelos, Mexico
	\label{inst3}
}

\date{Received; Accepted}

\abstract
{Embedded planets disturb the density structure of the ambient disk, and gravitational back-reaction possibly will induce
 a change in the planet's orbital elements. Low-mass planets only have a weak impact on the disk, so their wake's torque can be
 treated in linear theory. Larger planets will begin to open up a gap in the disk through nonlinear interaction.
 Accurate determination of the forces acting on the planet requires careful numerical analysis. 
 Recently, the validity of the often used fast orbital advection algorithm (FARGO) has been put into question, and special
 numerical resolution and stability requirements have been suggested.  
}
{We study the process of planet-disk interaction for low-mass planets of a few Earth masses, and
 reanalyze the numerical requirements to obtain converged and stable results. One focus lies on the
 applicability of the FARGO-algorithm. Additionally, we study the difference of two and three-dimensional simulations,
 compare global with local setups, as well as isothermal and adiabatic conditions.
 }
{We study the influence of the planet on the disk through two- and three-dimensional hydrodynamical
 simulations. To strengthen our conclusions we perform a detailed numerical comparison where several upwind and Riemann-solver
 based codes are used with and without the FARGO-algorithm.
}
{With respect to the wake structure and the torque density acting on the planet, we demonstrate that the FARGO-algorithm 
 yields correct a correct and stable evolution for the planet-disk problem, and that at a fraction of the regular cpu-time.
 We find that the resolution requirements for achieving convergent results in unshocked regions are rather modest and depend on the
 pressure scale height $H$ of the disk.
 By comparing the torque densities of two- and three-dimensional simulations we show that a suitable vertical averaging procedure for the force gives an
 excellent agreement between the two. We show that isothermal and adiabatic runs can differ considerably, even for adiabatic indices
 very close to unity.
}
{}

\keywords{	accretion, accretion disks --
			protoplanetary disks --
			hydrodynamics --
			methods: numerical --
			planets and satellites: formation
          }
\maketitle

\section{Introduction}
\label{sec:introduction}

Very young planets that are still embedded in the protoplanetary disk will disturb the ambient density by their
gravity. This will lead to gravitational torques that can alter the orbital elements of the planet. For
massive enough planets, the wake becomes nonlinear, and gap formation sets in. 
In numerical simulations of embedded planets, different length scales have to be resolved, in particular when studying
low-mass planets. On the one hand, the global structure has to be resolved to be able to obtain the correct structure of the
wakes, i.e. the spiral arms generated by the planet, which requires a sufficiently large radial domain.
The libration of co-orbital material on horseshoe streamlines requires a full azimuthal extent of 2 $\pi$ radians
to be properly captured.
On the other hand, the direct vicinity of the planet has
to be resolved to study detail effects, such as horseshoe drag or accretion onto the planet.

To ease computational requirements, often planet-disk simulations are performed in the two-dimensional (2D) thin disk
approximation, because a full three-dimensional (3D) treatment with high resolution is still very time-consuming.
However, even under this reduced dimensionality, the problem is still computationally very demanding.
The main reason is the strongly varying timestep size caused by the differentially rotating disk. Because the disk
is highly supersonic with (azimuthal) Mach numbers of about 10 to 50, the angular velocity at the inner disk
will limit the timestep of the whole simulation, even though the planet or other regions of interest are located much farther out.
Changing to a rotating coordinate system will not help too much owing to the strong differential shear.
To solve this particular problem and speed up the computation, \citet{2000A&AS..141..165M} has developed 
a fast orbital advection algorithm (FARGO). This method consists of an analytic, exact shift in the hydrodynamical quantities
by approximately the average azimuthal velocity. The transport step utilizes only the residual velocity, which is
close to the local sound speed. Depending on the grid layout and the chosen radial range, a very large speed-up
can be achieved, while at the same time the intrinsic numerical diffusion of the scheme is highly reduced 
\citep{2000A&AS..141..165M,2000ASPC..219...75M}.

The original version of the algorithm has been implemented into the public code {\tt FARGO},
which is very often used in planet-disk and related simulations. The accuracy of the FARGO-algorithm has been demonstrated in
a detailed planet-disk comparison project utilizing embedded Neptune and Jupiter mass planets \citep{2006MNRAS.370..529D}.
There, it has been shown that it leads to identical density profiles near the planet and total torques acting on the planet.
Meanwhile, similar orbital advection algorithms have been implemented into a variety of different codes in two and three spatial
dimensions, e.g. {\tt NIRVANA} \citep{1997CoPhC.101...54Z,2009A&A...506..971K},
{\tt ATHENA} \citep{2008JCoPh.227.4123G,2010ApJS..189..142S},
and {\tt PLUTO} \citep{2007ApJS..170..228M, Pluto2012}.
Despite these widespread applications, it has been claimed recently that usage of the FARGO-algorithm (here in  connection with {\tt ATHENA})
may lead to an unsteady behavior of the flow near the planet,
which even affects the wake structure of the flow farther away from the planet \citep{2011ApJ...741...56D}.

In the same paper, \citet{2011ApJ...741...56D} note that a very high numerical grid resolution
is required to obtain a resolved flow near the planet. In particular, they analyze the smooth,
unshocked wake structure close to the planet and infer that a minimum spatial resolution of about 256 gridcells per
scale height $H$, of the disk is needed to obtain good agreement with linear studies.
New simulations with a moving mesh technique also seem to indicate the necessity of very high resolutions
\citep{2012ApJ...755....7D}. 

Because a robust, fast, and reliable solution technique is mandatory in these type of simulations, we decided to address
the planet-disk problem for a well defined standard setup, which is very close to the one used in \citet{2011ApJ...741...56D}.
To answer the question of the validity of the FARGO-algorithm and estimate the resolution requirements, we applied several different
codes to an identical problem. These range from classical second-order upwind schemes (e.g. {\tt RH2D}, {\tt FARGO}) to modern Riemann-solvers
such as {\tt PLUTO}. The characteristics of these codes are specified in Appendix \ref{seca:codes}. 

Another critical issue in planet-disk simulations is the selection of a realistic treatment of the gravitational
force between the disk and the planet. Because the planet is typically treated as a pointmass and located within the
numerical grid, regularization of the potential is required. In addition, physical smoothing is required 
to account for the otherwise neglected vertical thickness of the disk. The magnitude of this smoothing is highly relevant,
since it influences the torques acting on the planet \citep{2002A&A...387..605M}, and the smoothing parameter has even entered analytical
torque formulas \citep{2009ApJ...703..857M,2010MNRAS.401.1950P}. Because the 2D equations are obtained by a vertical averaging procedure,
the force should be calculated by a suitable vertical integration as well. This approach has been undertaken recently by
\citet{2012A&A...541A.123M}, who show that the smoothing length is indeed determined by the vertical thickness of the disk, and is
roughly on the order of $0.7\,H$.
They show in addition that the change of the disk thickness induced by the presence of the planet has to be taken into
account. Because in recent simulations very short smoothing lengths have been used in 2D simulations \citep{2011ApJ...741...56D,2012ApJ...755....7D},
we compare our 2D results on the standard problem to an equivalent 3D setup and infer the required right amount of smoothing.

Finally, we performed additional simulations for different equations of state. The first set of simulations deals with 
the often used locally isothermal setup, while in comparison simulations we explore the outcome of adiabatic runs. This
is important because some codes may not allow for treating an isothermal equation of state.
Here, we use different values for the ratio of specific heat $\gamma$. In particular, a value of $\gamma$ very close to unity
has often been quoted as closely resembling the isothermal case. We show that this statement can depend on the physical problem.
In particular, in flows where the conservation of the entropy along streamlines is relevant, there can be strong
differences between an isothermal and an adiabatic flow, regardless of the value chosen for $\gamma$. 
For the planet-disk problem this has already been shown by \citet{2008A&A...478..245P}.

In the following section \ref{sec:model} we describe the physical and numerical setup of our standard model, and present
the numerical results in Sect.~\ref{sec:std-results}.
The validity of the FARGO-algorithm is checked in Sect.~\ref{sec:numerics}.
Alternative setups (nearly local, 2D versus 3D, adiabatic) are discussed in Sect.~\ref{sec:alter}. 
The transition of the wake into a shock front is discussed in Sect.~\ref{sec:shock}, and in the last section, we
summarize our results.
\section{The physical setup}
\label{sec:model}
We study planet-disk interaction for planets of very low masses that are embedded in a protoplanetary disk.
Most of our results shown refer to two-dimensional (2D) simulations, using the vertically integrated hydrodynamic
equations.  For validation and comparison purposes some additional full three-dimensional (3D) models have been performed
using a similar physical setup.
In all cases, we assume that the disk lies in the $z=0$ plane and use, for the 2D models, a cylindrical
coordinate system ($r, \varphi, z$), while in the 3D case we use spherical polar coordinates  ($R, \varphi, \theta$).

We consider locally isothermal as well as adiabatic models.
In the first case, the thermal structure of the disk is kept fixed and we chose for the standard model a constant aspect ratio, $h = H/r$.
Here $r$ is the distance to the star and $H$ is the local vertical scale height of the disk
\begin{equation}
\label{eq:height}
	H = \frac{c_\mathrm{s}}{\Omega_\mathrm{K}},
\end{equation}
where $c_{\rm s}$ is the isothermal sound speed and ${\Omega_{\rm K}} = (G M_*/r^3)^{1/2}$ is the Keplerian angular velocity
around the star.
During the computations the orbital elements of the planet remain fixed at their initial values, i.e. we 
assume no gravitational back-reaction of the disk on the planet or the star. This allows to formulate the problem scale free
in dimensionless units. The planet, whose mass is specified in term of its mass ratio $q = M_{\rm p}/M_*$,
is placed on a circular orbit at the distance $r=1$ and has angular velocity $\Omega_{\rm p} =1$,
i.e. one planetary orbit in these units is $2 \pi$.
The initial surface density $\Sigma_0$ is constant and can be chosen arbitrarily as it scales out of the equations.

In the 2D case the basic equations for the flow in the $r-\phi$ plane are given by equation of continuity 
\begin{equation}
	\label{eq:continuity}
	\frac{\partial \Sigma}{\partial t}  +  \nabla \cdot ( \Sigma \vec{u})  = 0,
\end{equation}
the momentum equation
\begin{equation}
	\label{eq:momentum}
	\frac{\partial \Sigma \vec{u}}{\partial t}  +  \nabla \cdot ( \Sigma  \vec{u} \, \vec{u}) \, = \, - \nabla P \, - \,  \Sigma  F_{\rm ext} \,,
\end{equation}
and the equation of energy
\begin{equation}
	\label{eq:energy}
	\frac{\partial e}{\partial t}  +  \nabla \cdot ( e  \vec{u} ) \, = \, - P \, \nabla \cdot  \vec{u} \,.
\end{equation}
Here, $e$ is the energy density (energy per surface area), and $P = (\gamma -1)e$ denotes the vertically integrated pressure.
In the isothermal case the energy equation is not evolved and the pressure reduces to $P = \Sigma c_\mathrm{s}^2$, where
$c_s (r)$ is a given function.
The external force 
\begin{equation}
          F_\mathrm{ext} = F_*  +  F_\mathrm{p} + F_\mathrm{inertial}.
\end{equation} 
contains the gravitational specific forces (accelerations) exerted by the star,
the planet and the inertial specific forces due to the accelerated and rotating coordinate system.

For the gravitational force generated by the central star and the inertial part of the force we use standard expressions.
The planetary force is more crucial because it influences for example the magnitude of the torque generated by the planet.
In our 2D standard model we derive it from a smoothed potential and we use the following very common form 
\begin{equation}
        \label{eq:planet_2dpot}
        \Psi^\mathrm{2D}_\mathrm{p} = - \frac{G M_\mathrm{p}}{\left(s^2 + \epsilon_\mathrm{p}^2\right)^\frac{1}{2}},
\end{equation}
where $s$ is the distance from the planet to the grid-point under consideration, 
and $\epsilon_\mathrm{p}$ is the smoothing length to the otherwise
point mass potential. It is introduced to avoid numerical problems at the location of
the planet. Then, $F_\mathrm{p} = - \nabla \Psi^\mathrm{2D}_\mathrm{p}$.
Alternatively, we will use in the 2D simulations a vertically averaged version of $F_\mathrm{p}$, following
\citet{2012A&A...541A.123M}, which we outline in more detail below.

Even though we use a non-zero but very small viscosity, we do not specify those terms explicitly in the above equations, see for example
\citet{1999MNRAS.303..696K}. The viscosity is so small, that it does not influence the flow on the relevant scales but is just included
to enhance stability and smoothness of the flow. 
In the inviscid case, the dynamical evolution of the system is controlled by the planet to star mass-ratio $q$ and the pressure
scale height $H$. A dimensional analysis by \citet{1996ApJS..105..181K} has shown that the relevant quantity is the nonlinearity parameter
\begin{equation}
        {\cal M}  =  \frac{q^{1/3}}{h}.
\end{equation}

\subsection{The standard model}
\label{subsec:standard}
\begin{table}
	\caption{
		\label{tab:standard}
		The physical and numerical parameter for the 2D standard model, which
		consists of a locally isothermal, two-dimensional disk with an embedded planet.
	}
	\centering
	\renewcommand\arraystretch{1.2}
\begin{tabular}{lll}
  Parameter &  Symbol  &   Value  \\
\hline
  mass ratio    &   $q = M_{\rm p}/M_*$  &  $6 \times 10^{-6}$   \\
  aspect ratio  &    $h= H/r$  &   $0.05$  \\
  nonlinearity parameter   &    ${\cal M}=q^{1/3}/h$  &   $0.36$    \\
\hline
  kinematic viscosity  &    $\nu$  &   $10^{-8}$    \\
  potential smoothing  &  $\epsilon_\mathrm{p}$  &    0.1 H \\
  radial range  &     $r_\mathrm{min}$ -- $r_\mathrm{max}$  &  $0.6$ -- $1.4$ \\
  angular range  &     $\phi_\mathrm{min}$ -- $\phi_\mathrm{max}$  &  $0$ -- $2\pi$ \\
  number of grid-cells    &  $N_r \times N_\phi $          &  $256 \times 2004$ \\
  spatial resolution     &  $\Delta r$          &  $H/16$  \\
  damping range at $r_\mathrm{min}$ & &  $0.6$ -- $0.7$  \\
  damping range at $r_\mathrm{max}$ & &  $1.3$ -- $1.4$  \\
\hline
\end{tabular}
\end{table}

The standard model refers to a 2D disk with a locally isothermal setup,
where the temperature is a given function of radius, that does not evolve in time. 
The parameters for our standard model are specified in Table~\ref{tab:standard}. The first two quantities, the mass and the aspect ratio,
determine the problem physically. As we intend to model the linear case in the standard model, we chose a small planet
with $q = 6 \times 10^{-6}$ which refers to a $2 M_\mathrm{Earth}$ planet for a solar mass star. For the disk's thickness
we assume $h = 0.05$.
For later purposes we list the nonlinearity parameter in the third row, which is $0.36$ here.
All results are a function of $q$ and $H$ only, if we assume a vanishing viscosity.
In the present situation, where we model indeed the inviscid case, we have chosen nevertheless a small non-zero kinematic viscosity $\nu = 10^{-8}$,
given in units of $a_p^2 \, \Omega_K(a_p)$. This is equivalent to an $\alpha$-value of $4 \times 10^{-6}$ for a $H/r=0.05$ disk,
a value that is considered to be much smaller than even a purely hydrodynamic viscosity in the disk. We have opted for the very small non-zero
value for numerical purposes. 
For the planetary gravitational potential $\Psi^\mathrm{2D}_\mathrm{p}$ we chose a value of $\epsilon = 0.1 H$ in the standard model.
We have selected this small value for $\epsilon$ primarily for comparison reasons to make contact to the recent simulations of 
\citet{2011ApJ...741...56D,2011ApJ...741...57D}, who suggest a very small smoothing length. Below we will demonstrate that
for physical reasons a much larger smoothing is required, which can be obtained by a suitable vertical averaging procedure
\citep{2012A&A...541A.123M}.

A whole annulus is modelled with a radial range from $r_\mathrm{min}= 0.6$ to $r_\mathrm{max}=1.4$.
We have chosen this domain size to capture all of the torque producing region,
which has a radial range of typically a few vertical scale heights, see e.g. \citet{2010ApJ...724..730D}.
The computational domain is covered by an equidistant grid which has $256\times 2004$ grid-cells. This results
in a resolution of $H/16$ at the location of the planet for the standard model.
To reduce or even avoid reflection from the inner and outer boundary we apply a damping procedure where,
within a specified radial range, all dynamical variables are damped towards their initial values.
Specifically, we use the prescription described in \citet{2006MNRAS.370..529D}, and write 
\begin{equation}
    \frac{d X}{d t}  =  - \frac{X(t) - X_0}{\tau} \, R(r) \,,
\end{equation}
where $X \in \{\Sigma, u_r\}$, and $R(r)$ is a ramping function increasing quadratically from zero to unity
(at the actual boundary) within the radial damping regions. The relaxation time $\tau$ is given by a fraction of the
orbital periods $T_\mathrm{orb}$ at  $r_\mathrm{min}$ and $r_\mathrm{max}$, respectively.
Here, we use a value of $\tau = 0.03 T_\mathrm{orb}$. For more details on the procedure see
\citet{2006MNRAS.370..529D}. We note, that it is sufficient to damp only the radial velocity $u_r$ (plus
$u_\theta$ in 3D simulations) which may be useful
in radiative simulations where the density stratification may not be know a priori.   
In test simulations (not displayed here) that use a larger radial range we have found identical results
concerning the torque and wake structure induced by the planet.

\subsection{Initial setup and boundary conditions}
\label{subsec:initial}
The initialization of the variables $\Sigma, T, u_r, u_\phi$ is chosen such that without the planet the
system would be in an equilibrium state. Here, we chose a constant $\Sigma_0$, and a temperature gradient
such that the aspect ratio $h = H/r$ is constant. That results in $T(r) \propto  r^{-1}$, which is fixed for the
(locally) isothermal models.
The radial velocity is zero initially, $u_r =0$, and
for $u_\phi$ we assume a nearly Keplerian azimuthal flow, corrected by the pressure gradient
\begin{equation}
    u_\phi(t=0)  =  r \Omega_\mathrm{K} \, ( 1 - h^2 )^{1/2} \,.
\end{equation}
The planet with mass ratio $q$ is placed at $r=1$ and $\phi = \pi$, i.e. in the middle of the computational domain. 
For some models the gravitational potential of the planet is slowly switched on within the first 5 orbital periods, while
others do not use this ramping procedure for the potential. 
For small mass planets, the results that are typically evaluated at $30\,T_\mathrm{orb}$ (i.e. $t = 30 \, \cdot 2 \pi$), and
there is no difference between these two options.

Near the inner and outer radial boundaries the solution is damped towards the initial state, using the procedure as described above.
In addition we use reflecting boundaries directly at $r_\mathrm{min}$ and $r_\mathrm{max}$.
In the azimuthal direction we use periodic boundaries.

\subsection{Numerical methods and codes}
\label{subsec:numerical}
Because one goal of this paper is to verify the accuracy of numerical methods, we apply several different codes to this
physical problem.
The 2D case is run using the following codes {\tt FARGO}, {\tt NIRVANA}, {\tt RH2D}, and {\tt PLUTO}.
All of these are finite volume codes utilizing a second order spatial discretization.
Additionally, all are empowered with the orbital advection speed-up, known as the FARGO-algorithm, as developed by \citet{2000A&AS..141..165M},
but can be used without this algorithm as well. The first 3 codes in the list have been used and described in \citet{2006MNRAS.370..529D}.
The last code, {\tt PLUTO}, is a multi-dimensional Riemann-solver based code for magnetohydrodynamical flows \citep{2007ApJS..170..228M},
which has been empowered recently with the  FARGO-algorithm \citep{Pluto2012}.

In the standard 2D-setup the simulations are performed in a cylindrical coordinate system that co-rotates with the embedded planet, and
the star is located at the origin. This implies that inertial forces such as Coriolis and centrifugal force as well as an
acceleration term to compensate for the motion of the star have to be included in the external force term $F_\mathrm{ext}$.
Additionally, the FARGO-algorithm is applied which leads to a speedup of around 10 for the standard model, and possibly even more
for the higher resolution models. For testing purposes the FARGO-algorithm can be alternatively switched on or off.

The 3D models are run in spherical polar coordinates. For reference we have quoted the inertial terms and their conservative
treatment in Appendix \ref{seca:eqn-rotframe}.
These are run using the following codes {\tt FARGO3D}, {\tt NIRVANA}, and {\tt PLUTO}, where  {\tt FARGO3D} is a newly developed
3D extension of {\tt FARGO}. For a description see Appendix \ref{seca:codes}.
For the timestep we typically use 0.5 of the Courant number (CFL). 
In Sect.~\ref{sec:numerics} we describe the outcome of the comparison in more detail.

\section{Results for the standard case}
\label{sec:std-results}

\begin{figure}
        \centering
        \includegraphics[width=\columnwidth]{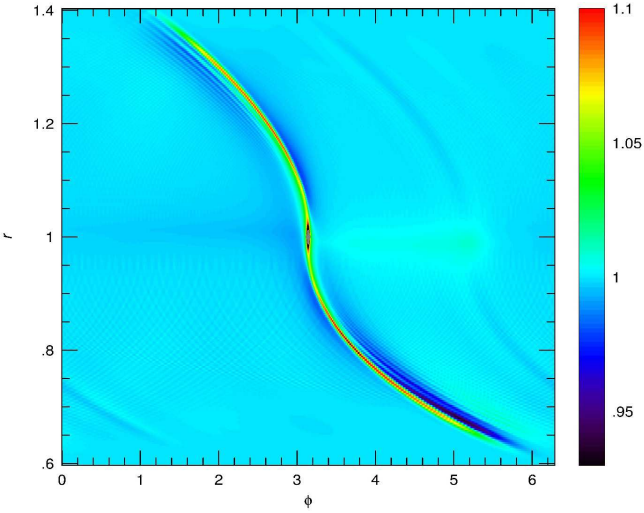}
        \caption{
                \label{fig:spiral-std}
                Density structure of the 2D standard model as generated by an embedded planet with $q=6 \times 10^{-6}$ in a disk
               having the aspect ratio $H/r = 0.05$. Shown is the configuration after $30\,T_\mathrm{orb}$. The density is scaled
              linearly. 
        }
\end{figure}

To set the stage and illustrate the important physical effects, we first present the results of our simulations
for the 2D standard case using the parameters according to Table \ref{tab:standard}. For the simulations
in this section we used the {\tt RH2D} code unless otherwise stated.
Below, we will then discuss the variations from the standard model. 

After the insertion of the planet, the planet's gravitational disturbance generates two wakes, in the form of trailing spiral arms.
The basic structure of the surface density, $\Sigma$, is shown in Fig.~\ref{fig:spiral-std}. 
As seen from the plot, the used damping procedure ensures that reflections by the radial boundaries are minimized.
There is indication of vortex formation as can be seen by the additional structure in the right hand side of Fig.~\ref{fig:spiral-std}.
Vortices induced by the planet occur for low viscosity disks and have been seen already in earlier simulations 
\citep[see e.g.][]{2005ApJ...624.1003L,2006MNRAS.370..529D,2009ApJ...690L..52L}. Here, we will not 
discuss this issue any further.

The relevant quantity to study the physical consequences of the interaction of the embedded planet with the ambient disk
is the gravitational torque exerted on the planet by the disk.
For that purpose it is very convenient to calculate the radial torque distribution per unit disk mass,
$d \Gamma(r)/dm$, which we define here, following \citet{2010ApJ...724..730D}, through the definition of the total
torque, $\Gamma_{\rm tot}$, acting on the planet
\begin{equation}
\label{eq:gamm-tot}
            \Gamma_{\rm tot} = 2 \pi \int \frac{d \Gamma}{d m} (r) \, \Sigma(r) \, r dr \,.
\end{equation}
Here, $d \Gamma(r)$ is the torque
exerted on the planet by a disk annulus of width $dr$ located at the radius $r$ and having the mass $dm$.
As $d \Gamma(r)/dm$ scales with the mass ratio squared and as $(H/r)^{-4}$, we rescale our results accordingly in units of
\begin{equation}
\label{eq:gamm0}
     \left( \frac{d \Gamma}{d m}\right)_0 =  \Omega_\mathrm{p}^2(a_\mathrm{p}) a_\mathrm{p}^2  q^2 \left(\frac{H}{a_\mathrm{p}}\right)^{-4},
\end{equation}
where the index $p$ denotes that the quantities are evaluated at the location of the planet, which has the semi-major axis $a_\mathrm{p}$.
\begin{figure}
        \centering
        \includegraphics[width=\columnwidth]{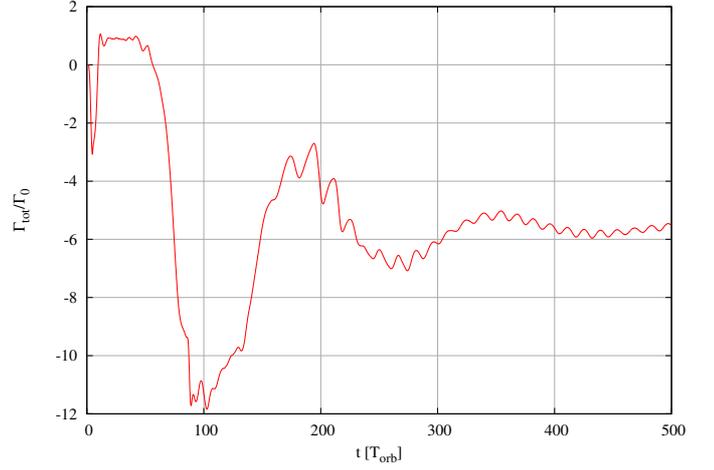}
        \caption{
                \label{fig:tx0-std}
                The total torque, $\Gamma_{\rm tot}$, in units of $\Gamma_0$ (see Eq.~\ref{eq:gamtot0}),
                 acting on the planet vs. time for the 2D standard model.
                Shortly after insertion, the torque is positive and approximately constant
                between 10 and 40 orbits. At later times it saturates  due to
                mixing of the material within the horseshoe region.
        }
\end{figure}
\begin{figure}
        \centering
        \includegraphics[width=\columnwidth]{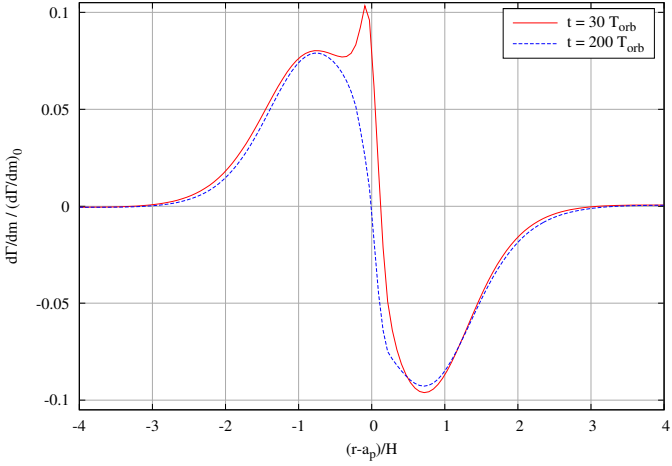}
        \caption{
                \label{fig:gam0-std}
                The radial torque density in units of $(d\Gamma/d m)_0$ (see Eq.~\ref{eq:gamm0})
               at 30 and 200 $T_\mathrm{orb}$ for the 2D standard model.
                The torque enhancement and spike near $r=1$ at $t=30\,T_\mathrm{orb}$ is due to the unsaturated corotation torque.
                At later times, here shown at $t=200\,T_\mathrm{orb}$, only the Lindblad contributions due to the spiral arms remain.
        }
\end{figure}
The time evolution of the total torque, $\Gamma_{\rm tot}$, is displayed in Fig.~\ref{fig:tx0-std} for the first 500 orbits. 
The total torque is stated in units of
\begin{equation}
\label{eq:gamtot0}
     \Gamma_0 =  \Sigma_0 \, \Omega_\mathrm{p}^2(a_\mathrm{p}) a_\mathrm{p}^4  q^2 \left(\frac{H}{a_\mathrm{p}}\right)^{-2} \,.
\end{equation}
In this simulation, the planetary potential has been ramped up during the first 5 orbits. After insertion of the 
planet the total torque becomes first positive and remains constant at this level for about 30 orbits. In this phase the co-orbital torque,
in particular the horseshoe drag, is fully unsaturated and gives rise to a total positive torque. 
In our situation of an isothermal disk this comes about
because of a non-vanishing vortensity gradient across the horseshoe region which generates in this case a strong positive
corotation torque \citep{1979ApJ...233..857G}.
The vortensity, which is defined as vorticity divided by surface density, is here given by 
\begin{equation}
    \zeta =  \frac{(\nabla \times \vec{u})|_z}{\Sigma} \,.
\end{equation}
As can be seen from this definition, for a constant $\Sigma$ disk the radial gradient of $\zeta$ is $\propto r^{-3/2}$,
which leads to the strong vortensity-related torque.
However, due to the different libration speeds, the material within the corotation region mixes and the gradients of
potential vorticity and entropy are wiped out in the absence of viscosity
\citep{2001MNRAS.326..833B,2001ApJ...558..453M}. Consequently, the torques drop again, and oscillate on timescales 
on the order of the libration time towards a negative equilibrium value, which is given by the Lindblad torques
generated by the spiral arms. This saturation of the vortensity-related torque related torque has been analyzed for example by
\citet{2007LPI....38.2289W} through an analyzes of streamlines within the horseshoe region for an inviscid disk,
and  later through two-dimensional hydrodynamic simulations by \citet{2010ApJ...723.1393M,2011MNRAS.410..293P}.
The strength of the (positive) corotation torque depends strongly on the smoothing of the gravitational potential.
 For the chosen small $\epsilon =0.1$ this results in a positive total torque. For more realistic values of $\epsilon \approx 0.6-0.7$,
 $\Gamma_{\rm tot}$ will usually be negative, see Section~\ref{subsec:2d3d}.

To study the spatial origin of the torques we analyze the radial torque density for the standard model. 
In Fig.~\ref{fig:gam0-std} the torque density $d\Gamma/dm$, according to Eq.~\ref{eq:gamm0} is displayed vs. radius in units
of $(d\Gamma/d m)_0$. Two snapshots are displayed, one at $t=30\,T_\mathrm{orb}$ where the torque is fully unsaturated and one at
$t=200\,T_\mathrm{orb}$ where the torque is saturated. 
We note, that for the torque calculation we use a tapering function near the planet to avoid contributions
of material that is bound to the planet, or is so close to yield large torque fluctuations due to numerical discretization
effects.
We use the form as given in \citet{2008A&A...483..325C} which reads 
\begin{equation}
\label{eq:taper}
   f(s)= \left[ \exp\left( - \, \frac{ \, s-r_\mathrm{t}}{0.1 r_\mathrm{t}} \, \right)+1\right]^{-1} \,,
\end{equation}
with a tapering length of $r_\mathrm{t} = 0.8\,R_\mathrm{H}$.
Here, $R_\mathrm{H} = (q/3)^{1/3} a_\mathrm{p}$ is the Hill radius of the planet. 
Such a tapering is particularly useful for massive planets that form a disk around them \citep{2009A&A...502..679C}.
Around lower mass planets, with ${\cal M} \lesssim 0.6$, circumplanetary disks do not form \citep{2006ApJ...652..730M} and a large
tapering is not required. Indeed, we found that for values of $r_\mathrm{t}$ in the range of $0.4-1.0 R_\mathrm{H}$
there is not a large difference in the measured torques in equilibrium. For example, the variations of the total torque
in Fig.~\ref{fig:tx0-std} are less than $5\%$. 

The torque density in the fully saturated phase, at $t=200\,T_\mathrm{orb}$ (Fig.~\ref{fig:gam0-std}, blue line), is positive inside of the planet and negative
outside of the planet. The positive contribution of this Lindblad torque comes from the inner spiral arm,
and the negative part from the outer one.
The distribution at the earlier time, $t=30\,T_\mathrm{orb}$, shows an additional contribution and spike just inside of the
planet. This part is due to the horseshoe drag which is subject to the described saturation process.   

\begin{figure}
        \centering
        \includegraphics[width=\columnwidth]{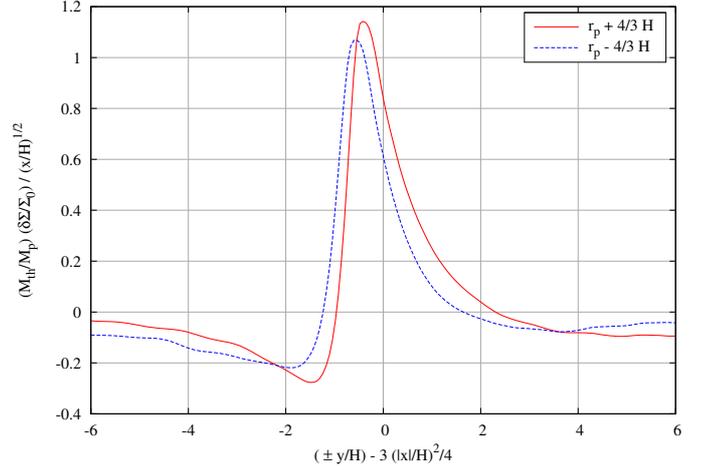}
        \caption{
                \label{fig:wake0-std}
                Normalized azimuthal density profile of the inner and outer wake at radial distances $\pm 4/3 H$ away from the planet
          at 30 $T_\mathrm{orb}$ for the isothermal 2D standard model. The coordinates $x$ and $y$ refer to local Cartesian coordinates, see
               Eq.~(\ref{eq:xy}). The 'plus' sign in the $x$-axis label refers to the blue curve at $r_\mathrm{p} -  4/3 H$,
              and the 'minus' sign
               to the red curve at $r_\mathrm{p} +  4/3 H$. The upstream side of the wake is to the left for both curves.
        }
\end{figure} 

To study the wake properties generated by the planet we use here a quasi-Cartesian local coordinate system centered
on the planet to allow direct comparison to previous linear results.
Specifically, we define
\begin{equation}
\label{eq:xy}
       x=(r-r_\mathrm{p})  \quad \mbox{and} \quad  y = (\phi - \phi_\mathrm{p}) r_\mathrm{p} \,.
\end{equation}
 
In Fig.~\ref{fig:wake0-std} the relative density perturbations for the inner and outer wakes are shown along the azimuth.
They are displayed at a radial distance of $x = \pm 4/3 H$ from the location of the planet. For the normalization of the perturbed density 
we define first the thermal mass of the planet
\begin{equation}
               M_\mathrm{th} = \left(\frac{c_s^3}{G \Omega}\right)_\mathrm{p} =  h^3 \, M_* \,,
\end{equation}
where the quantities have to be evaluated at the location of the planet.
Then, the ratio of the planet mass to the thermal mass is given by
\begin{equation}
               \frac{M_\mathrm{p}}{M_\mathrm{th}} = \frac{q}{h^3} = {\cal{M}}^3 \,.
\end{equation}
Now, we follow \citet{2011ApJ...741...56D} and scale $\delta \Sigma = \Sigma(\phi) - \Sigma_0$ by the planet mass (in units of $M_\mathrm{th}$)
and normalize by $x/H$.

Due to the radial temperature variation and the cylindrical geometry, the inner and outer wake differ in their appearance. 
However, the general shape and magnitude is very similar to the linear results that have been
obtained for the local shearing sheet model \citep[see][]{2001ApJ...552..793G,2011ApJ...741...56D}.
Differences in the amplitude are presumably due to a different normalization. 
Because our results (in all simulations and with all codes) are consistently by factor of 3/2 larger than
those of \citet{2011ApJ...741...56D} we suspect that they might have used the normalization of
$M_\mathrm{th}$ as given by \citet{2001ApJ...552..793G} which differs exactly by this factor.
At the displayed distance from the planet the wake is expected to be in the linear regime,
which results in a smooth maximum. 
For this reason we do not expect a large dependence on the numerical resolution.
In the following, we will use only the outer wake to check
for possible variations due to setup, numerical methods and resolution.

\section{Testing numerics}
\label{sec:numerics}

To validate our results, and demonstrate that the FARGO-algorithm yields accurate results,
we varied the numerical setup, and used several different codes on the same physical problem.
In this Appendix we describe our studies in more detail.

\begin{figure}
        \centering
        \includegraphics[width=\columnwidth]{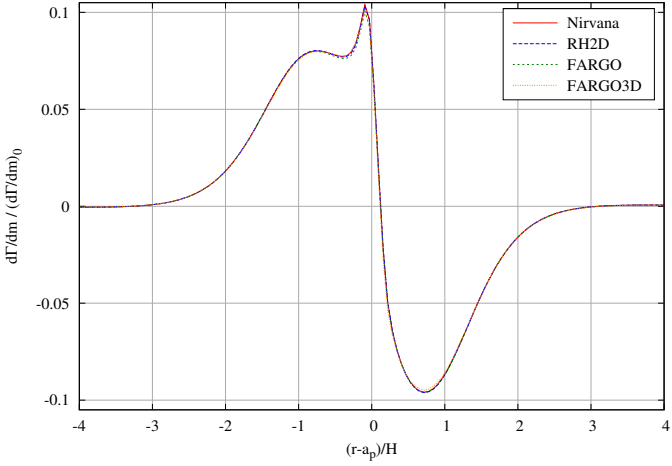}
        \caption{
                \label{figa:gamc-codes}
                The radial torque density of the 2D standard-problem in units of $(d\Gamma/d m)_0$
               at 30 $T_\mathrm{orb}$ for different codes at the standard resolution.
        }
\end{figure} 
\begin{figure}
        \centering
        \includegraphics[width=\columnwidth]{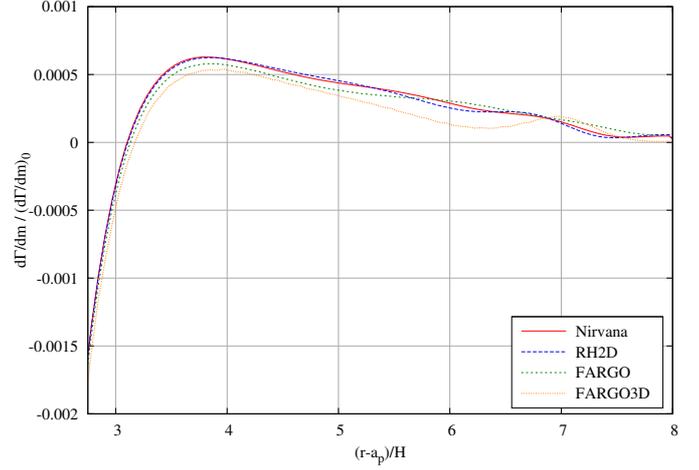}
        \caption{
                \label{figa:gamc2-codes}
                The radial torque density in units of $(d\Gamma/d m)_0$
               at 30 $T_\mathrm{orb}$ for the standard setup for various code.
               This is enlargement of Fig.~\ref{figa:gamc-codes}.
        }
\end{figure}
\begin{figure}
        \centering
        \includegraphics[width=\columnwidth]{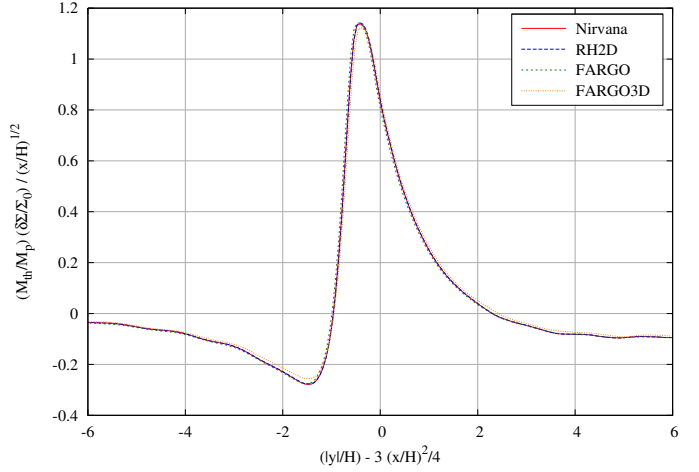}
        \caption{
                \label{figa:wakec-codes}
                Normalized azimuthal density profile of the outer wake at the radius $r_\mathrm{p} + 4/3 H$
               at 30 $T_\mathrm{orb}$ for  different codes at the standard resolution.
        }
\end{figure} 

\subsection{Using different codes on the 2D standard model}
\label{subsec:std-codes}
To support our findings on the torque density and wake form and demonstrate the accuracy of the used
codes, we ran the 2D standard model in the isothermal and the adiabatic version using all of the above codes.
All use the {FARGO}-setup and are run in the same (standard) resolution. 
The isothermal results for the torque density are shown in Figs.~\ref{figa:gamc-codes} and \ref{figa:gamc2-codes},
where the latter displays an enlargement of the first.  Clearly, the results agree extremely well for the different codes.
This includes the standard Lindblad torques as well a the detailed structure of the corotation torque.
The {\tt FARGO3D} code has been used in its 2D version for this test.

Recently, it has been shown by \citet{2011ApJ...741...56D,2012ApJ...747...24R} that the torque density 
$\Gamma(r)$ changes sign at a certain distance from the planet in contrast to the standard linear results
\citep{1979ApJ...233..857G}. Here, we show that this effect is reproduced in our simulations as well for all codes.
In Fig.~\ref{figa:gamc2-codes} we show that this reversal occurs at a distance $r_\pm \approx 3.1 H$ away from the
planet in good agreement with the linear results of \citet{2012ApJ...747...24R}.
Again, all codes agree well on this feature, only {\tt FARGO3D} shows small deviations. 

The corresponding wake form at $x = r_\mathrm{p} + 4 H$ is displayed in Fig.~\ref{figa:wakec-codes}. It is identical for all four cases,
which shows the consistency and accuracy of the results and codes.

\subsection{The FARGO treatment}
\label{subsec:std-fargo}
\begin{figure}
        \centering
        \includegraphics[width=\columnwidth]{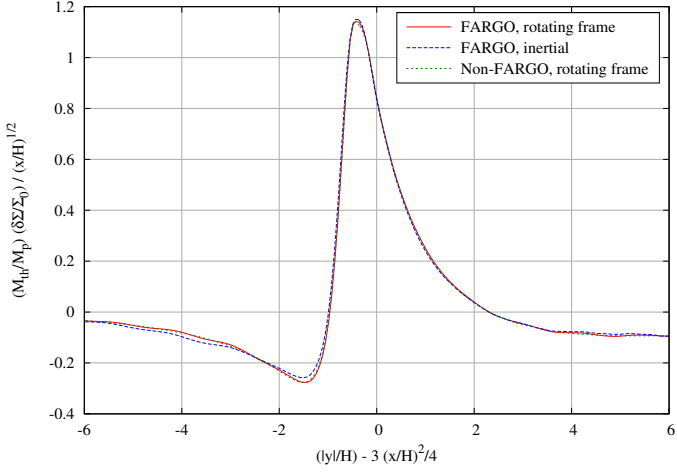}
        \caption{
                \label{figa:wakec-fargo}
                Normalized azimuthal density profile of the outer wake at the radius $r_\mathrm{p} + 4/3 H$
               at 30 $T_\mathrm{orb}$ for the code {\tt RH2D} using different timesteps and a non-rotating frame.
        }
\end{figure}
In Fig.~\ref{figa:wakec-fargo} the wake form is analyzed for three different numerical usages of the  FARGO-algorithm
on the 2D standard setup. The first, red curve, corresponds to the
standard reference case using a corotating coordinate system and the FARGO-algorithm. For the second, blue curve, the simulation
has been performed in the inertial frame and using FARGO. In the mechanism of the algorithm, the quantities in each ring
are first shifted according to the overall mean angular velocity of the ring,
and then advected using the residual velocity \citep{2000A&AS..141..165M}.
As a result, theoretically it should not matter whether the coordinate system is rotating or not.
This is exactly what we find in our simulations, as the blue curve is very similar to the red one. Small differences
can be produced by the planetary potential which is time dependent in the latter case, as the planet is moving, and lies
at different locations with respect to the numerical grid.
Also, the number of timesteps used until 30 $T_\mathrm{orb}$ are identical for two runs ($12,866$ steps).
The third, green curve, corresponds to a model in the corotating frame without using the FARGO-algorithm. 
Because of the small timestep size in this case, over ten times more timesteps had to be used in this case ($137,750$ steps).
Nevertheless, the wake form is identical. These runs indicate, that the FARGO-algorithm captures the physics of the system
correctly. At the same time it comes with a much larger timestep and hence much reduced computational cost.
This also applies to modern Riemann-solvers such as {\tt PLUTO}, as shown in section~\ref{subsec:adi}.
\begin{figure}
        \centering
        \includegraphics[width=\columnwidth]{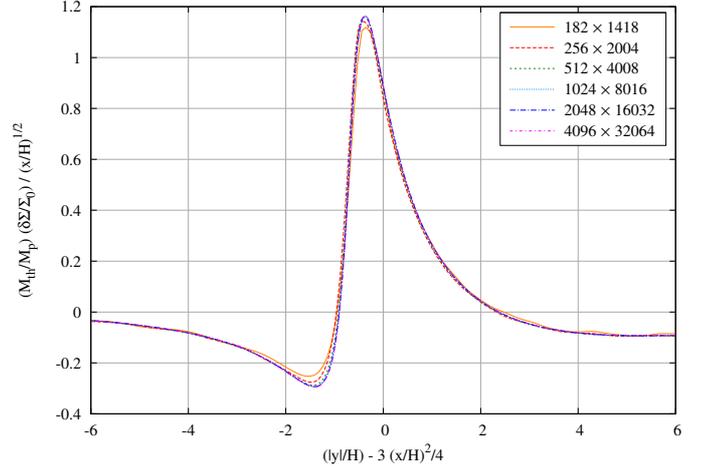}
        \caption{
                \label{figa:wakec-res}
                Normalized azimuthal density profile of the outer wake at the radius $r_\mathrm{p} + 4/3 H$
               at 30 $T_\mathrm{orb}$ using the {\tt FARGO}-code at different resolutions.
        }
\end{figure}
\subsection{Testing numerical resolution}
\label{subsec:std-resolution}
In order to estimate the effect of numerical resolution, we ran the 2D standard model using grid-sizes
ranging from $182 \times 1418$ all the way to $4096 \times 32064$. This is equivalent to grid resolutions
of $H/10$ to $H/256$. As shown in Fig.~\ref{figa:wakec-res}, the results are nearly identical at all
resolutions. The first two, lower resolution cases have a slightly lower trough just in front of the wake and a smaller
amplitude. As will be discussed later in Sect.~\ref{sec:shock}, the results for the different resolutions are so similar because
at this distance to the planet the wake is in the linear regime and has not steepened to a shock wave yet.
The resolution requirements at the shock front will be analyzed in Sect.~\ref{sec:shock}.

\begin{figure}
        \centering
        \includegraphics[width=\columnwidth]{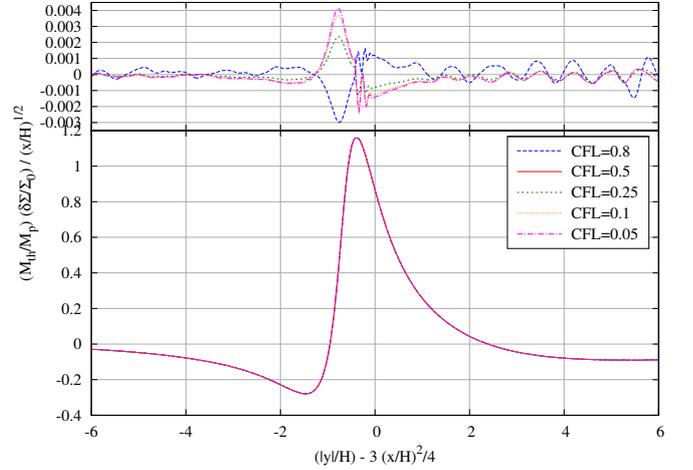}
        \caption{
                \label{fig:wake-cfl}
                Normalized azimuthal density profile of the outer wake at the radius $r_\mathrm{p} + 4/3 H$
               at 30 $T_\mathrm{orb}$ using the {\tt FARGO}-code with different Courant-numbers (CFL).
               The physical setup differs slightly from the standard problem and is described in Sect.~\ref{subsec:std-timestep}.
            In the upper panel the differences of the individual runs with respect to the standard, CFL=0.5, are displayed.
        }
\end{figure}

\subsection{Testing timestep and stability}
\label{subsec:std-timestep}
Finally, we would like to comment on possible timestep limitations due to the gravitational force
generated by the planet.
In our simulations we never found any unsteady evolution when using orbital advection. 
In contrast, the results of \citet{2011ApJ...741...56D} indicate an unsteady behaviour for larger
timesteps. They attribute possible instabilities to a violation of
an additional gravity related timestep criterion and advocate using very small timesteps, which would render
the FARGO-algorithm not applicable in very many cases.

To test specifically this statement we performed a suite of simulations on a setup very similar to that used
by \citet{2011ApJ...741...56D} in their Fig.~12. Due to the difficulty of {\tt RH2D} and {\tt FARGO} to use a
Cartesian local setup we used here a computational domain exactly as before with a grid-size of $1024\times 8016$ which gives
a resolution of 64 gridcells per scaleheight $H$. The planet mass is 1.33 $M_{\rm Earth}$, which is equivalent to a mass ratio
$q = 4 \times 10^{-6}$ or $M_{\rm p} = 3.2 \times 10^{-2} M_{\rm th}$. For the potential smoothing we choose $\epsilon = 0.08 H$
which is yields a planetary potential nearly identical to that of  \citet{2011ApJ...741...56D}. 
In Fig.~\ref{fig:wake-cfl} we display the results (using {\tt FARGO}) for different timestep sizes as indicated by the corresponding Courant-number.
The CFL = 0.5 case corresponds to our standard case. We made the timestep larger (CFL = 0.8) as well as smaller, down to CFL = 0.05. 
All cases yield identical results and do not show any sign of instability. In the upper panel the differences of the individual runs with respect to
the standard, CFL=0.5, are displayed. The performed runs with the {\tt RH2D}, {\tt FARGO} and {\tt FARGO3D} 
codes yield identical results, again with no signs of unsteady behaviour.
Here, {\tt FARGO3D} was run in the 2D version,
both with the setup as indicated above and with the local setup of Tab. 3, with resolution $h/64$. 
For all our runs, past the first two orbits the
wake profile at $x=1.33 H$ has achieved convergence to better than the 1\% level, regardless of the value of the
timestep size.
In Appendix~\ref{sec:time-gravity} we re-analyze possible stability requirements in the presence of gravity and find indeed stability
for the timestep sizes used with the FARGO-algorithm. 

Note, that for these runs we switched off the physical viscosity completely. We find that the result for the wake displayed in Fig.~\ref{fig:wake-cfl}
is in fact, due to the special scaling of the axes, identical to that of the standard problem as shown in the previous plots. 
Additionally, we have not seen any sign of unsteady behaviour. All of this indicates that our small value of the kinematic viscosity,
$\nu = 10^{-8}$ (in dimensionless units), is essentially negligible.

\section{Using alternative setups}
\label{sec:alter}
To illustrate how variations of individual properties of the standard model influence the outcome,
we performed additional simulations which are described in this section.
\begin{figure}
        \centering
        \includegraphics[width=\columnwidth]{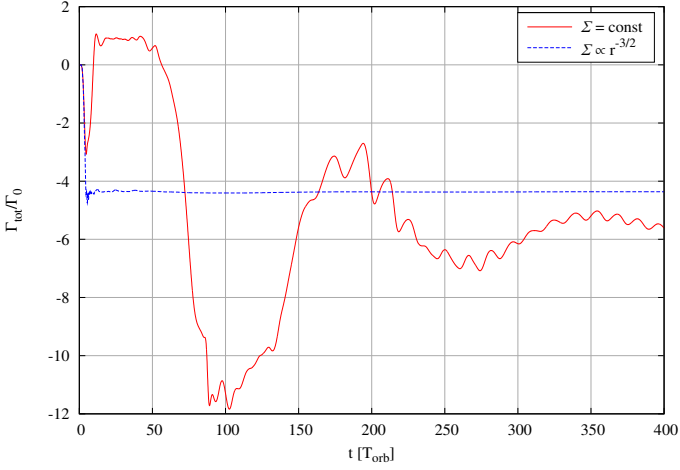}
        \caption{
                \label{fig:tx0-sigma}
                The total torque $\Gamma_{\rm tot}$, in units of $\Gamma_0$ (see Eq.~\ref{eq:gamtot0}),
                 acting on the planet vs. time. Shown are the 2D standard model (red),
                and a globally isothermal case (blue), with 
                 a different density profile, such that the potential vorticity gradient vanishes.
        }
\end{figure}

\subsection{Different radial stratification}
\label{subsec:slopes}
As shown above, in the initial evolution after embedding the planet the total torque is
positive due to a large positive horseshoe drag. The strength of this effect depends on the
radial gradients of potential vorticity, entropy (for simulations with energy equation), and
temperature \citep{2012arXiv1203.3294B}.
To minimize this effect we present an additional, alternative
setup where the gradients of potential vorticity (vortensity) and temperature vanish exactly.
Hence, we chose a setup with a density gradient $\Sigma \propto r^{-3/2}$ and $T = \mathrm{const}$.
The time evolution of the total torque for this model is displayed together with the standard case in
Fig.~\ref{fig:tx0-sigma}. Clearly, after the short switch-on period of the planet mass,
the total torque is negative and constant throughout the evolution.
This demonstrates that for this density profile, $\Sigma \propto r^{-3/2}$, which resembles (coincidently) the minimum mass solar nebula,
there is indeed no corotation torque present, and the flow settles directly to the Lindblad torque.
The final value for the total Lindblad torque differs slightly for the two models due to the different gradients in
density and temperature.
We note that for this setup, with vanishing vortensity gradient, there are also no vortices visible during the initial
evolution. 

\subsection{Comparing 2D and 3D simulations}
\label{subsec:2d3d}
\begin{table}
	\caption{
		\label{tab:std-3D}
		The numerical parameter for the 3D standard model
	}
	\centering
	\renewcommand\arraystretch{1.2}
\begin{tabular}{lll}
  Parameter &  Symbol  &   Value  \\
\hline
  smoothing radius  &   $r_\mathrm{sm}$  &  $0.25 R_\mathrm{H}$ \\
  radial range  &     $R_\mathrm{min}$ -- $R_\mathrm{max}$  &  $ 0.6$ -- $1.4$ \\
  angular range  &     $\phi_\mathrm{min}$ -- $\phi_\mathrm{max}$  &  $0$ -- $2\pi$ \\
  meridional range  &     $\theta_\mathrm{min}$ -- $\theta_\mathrm{max}$  &  $82\,^\circ$ -- $90\,^\circ$ \\
  number of grid-cells    &  $N_r \times N_\phi  \times N_\theta  $          &  $256 \times 2004 \times 39$ \\
  spatial resolution     &  $\Delta r$          &  $H/16$  \\
  damping range at $R_\mathrm{min}$ &  &  $0.6$ -- $0.7$  \\
  damping range at $R_\mathrm{max}$ & &  $1.3$ -- $1.4$  \\
\hline
\end{tabular}
\end{table}

\begin{figure}
        \centering
        \includegraphics[width=\columnwidth]{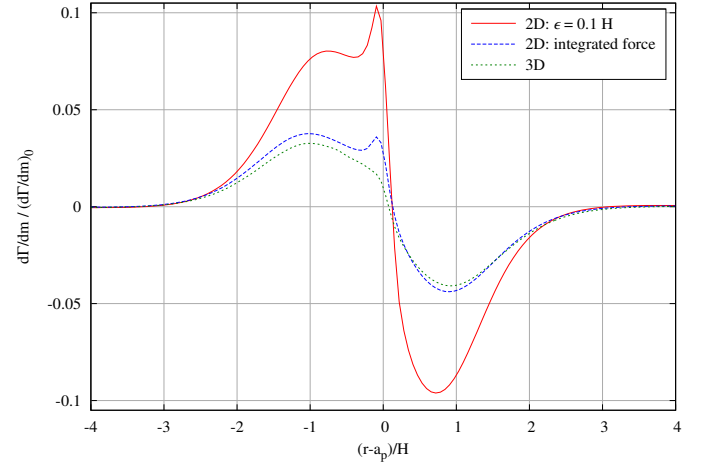}
        \caption{
                \label{fig:gamc-3D}
                The radial torque density in units of $(d\Gamma/d m)_0$
               at 30 $T_\mathrm{orb}$ for 2D and 3D simulations of the standard setup.
               The red curve corresponds to that in Fig.~\ref{fig:gam0-std}, the blue line to a 2D
                model with a vertical integrated gravitational force and the green to the
             3D model (using {\tt NIRVANA}).
        }
\end{figure}
\begin{figure}
        \centering
        \includegraphics[width=\columnwidth]{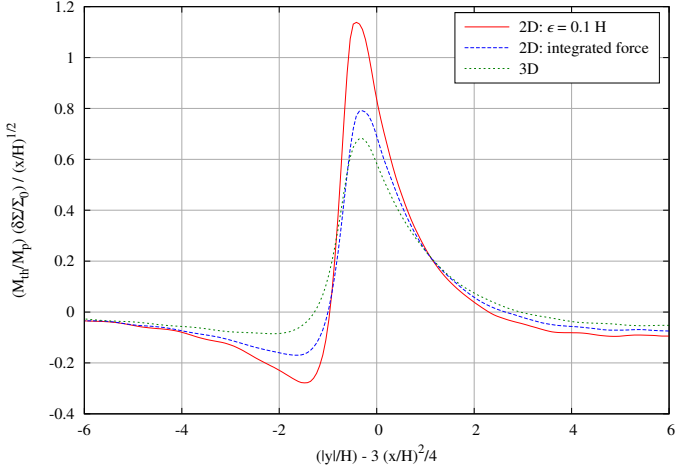}
        \caption{
                \label{fig:wakec-3D}
                Normalized azimuthal density profile of the outer wake at the radius $r_\mathrm{p} + 4/3 H$
               at 30 $T_\mathrm{orb}$  for 2D and 3D simulations of the standard setup.
              The color coding is identical to Fig.~\ref{fig:gamc-3D}.
        }
\end{figure}

The setup of the described standard case reduces the physical planet-disk problem to two dimensions.
However, even though the disk may be thin, corrections are nevertheless expected due its finite thickness. 
We investigate this by performing full 3D simulations using the same physical setup as in the standard 2D model.
The treatment of the inertial forces is outlined in Appendix \ref{seca:eqn-rotframe}. 
The additional numerical parameters are listed in Table~\ref{tab:std-3D}. The spatial extent and numerical resolution
is identical to the 2D model.
The initialization of the 3D density is chosen such that the surface density
is constant throughout. In the vertical direction the density profile is initialized with a Gaussian profile as expected
for vertically isothermal disks. The temperature is constant on cylinders. 
For the gravitational potential of the planet
we chose the so called cubic-form \citep{2009A&A...506..971K} which is exact outside a smoothing radius $r_\mathrm{sm}$ and smoothed
by a cubic polynomial inside of $r_\mathrm{sm}$. The advantage of this form lies in the fact that in 3D simulations the smoothing is required
only numerically and the cubic potential allows us to have the exact potential outside a specified radius, here $r_\mathrm{sm}$.
To calculate the torque the same tapering function (Eq.~\ref{eq:taper}) as before has been used. 

\begin{figure}
        \centering
        \includegraphics[width=\columnwidth]{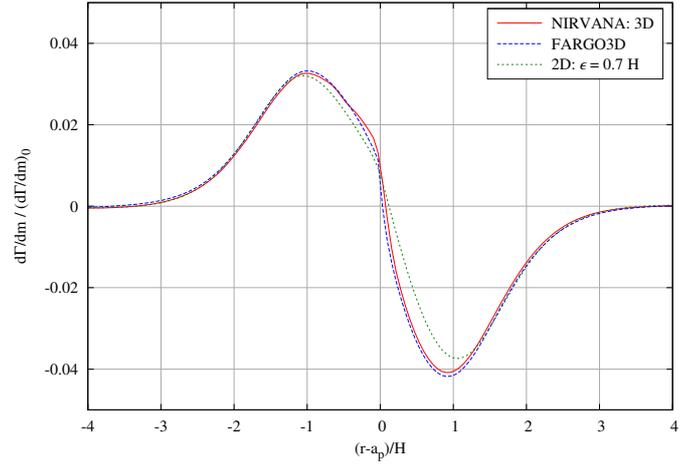}
        \caption{ 
                \label{fig:gamc-3Dx}
                The radial torque density in units of $(d\Gamma/d m)_0$
               at 30 $T_\mathrm{orb}$ for the 3D and 2D simulations of the standard setup.
               Compared are two 3D simulations (using {\tt NIRVANA} and {\tt FARGO3D}) with a
             2D simulation, using $\epsilon = 0.7$.
        }
\end{figure}

In  Fig.~\ref{fig:gamc-3D} we show the normalized torque density $d \Gamma/dm$ for 2D simulations in comparison to a full
3D simulation using the same physical setup. 
Due to the finite vertical extent the torques of the 3D model are substantially smaller
than for the corresponding 2D setup. 
As \citet{2012A&A...541A.123M} have shown recently, this discrepancy can be avoided by performing
a suitable vertical averaging procedure of the gravitational force. 
Specifically, the force acting on each disk element in a 2D simulation is calculated from the projected force that
acts in the midplane of the disk. Denoting the distance of the disk element to the planet with $s$, the force density
(force per area) is given by
\begin{equation}
        \label{eq:force_s}
        F_\mathrm{p}(s)
        =  - \int \rho \frac{\partial \Psi_\mathrm{p}}{\partial s} \,dz
        =  - G M_\mathrm{p} s \int \frac{\rho}{(s^2 + z^2)^\frac{3}{2}} \,dz \,,
\end{equation}
where $\Psi_\mathrm{p}$ is the physical 3D potential generated by the planet. For the vertical density stratification
a Gaussian density profile can be assumed for a vertically isothermal disk as a first approximation. 
However, the change of the vertical density as induced by the planet has to be taken into account.
The results using this averaging prescription in an approximate way \citep{2012A&A...541A.123M} is shown additionally
in Fig.~\ref{fig:gamc-3D} by the blue curve.
The overall behavior and magnitude is very similar to the full 3D results. For comparison,
a 2D model using a fixed $\epsilon=0.7 H$ (instead of $0.1 H$ of the standard model) for $\Psi^\mathrm{2D}_\mathrm{p}$
yields similar amplitude as the full 3D model but a slightly different shape (see  Fig.~\ref{fig:gamc-3Dx}).

In  Fig.~\ref{fig:wakec-3D} we show the corresponding wake profile for the 2D and 3D setup. 
For the 3D case, the surface density is obtained by integration along the $\theta$ direction at constant spherical radii, $R$.
Here, the wake amplitude of the
full 3D model is again reduced in contrast to the flat 2D case, with $\epsilon = 0.1$. The 2D model using the integrated
force algorithm yields here again a better agreement.
The 3D results displayed in these plots have been obtained with {\tt NIRVANA}, 
but usage of the new code {\tt FARGO3D} yields identical results, as is demonstrated in Fig.~\ref{fig:gamc-3Dx}.

These results demonstrate clearly, that the $\epsilon$-parameter in the 2D planetary potential ($\Psi^\mathrm{2D}_\mathrm{p}$) cannot be
chosen arbitrarily small, but has to be on the order of the scale height $H$ of the disk. Near to the planet a reduction is required
to account for the reduced thickness, see \citet{2012A&A...541A.123M}. Hence, the value of $\epsilon = 0.1\,H$ as chosen for the standard
setup is too small to yield good agreement with vertically stratified disks, and serves here only as a numerical illustration to connect
to previous linear and numerical results \citep{2001ApJ...552..793G,2011ApJ...741...56D}.
As shown by \citet{2012A&A...541A.123M} a value of $\epsilon = 0.7\,H$ yields similar amplitudes to the 3D case, in particular for the
Lindblad torque, see  Fig.~\ref{fig:gamc-3Dx}.
However, as can be seen from the figure, for the $\epsilon$-potential the relative strengths of the inner an outer torques
differ from the full 3D and the 2D vertically integrated case.

\subsection{Using a quasi-local setup}
\label{subsec:local}

To demonstrate the agreement of our simulations with previously published local results,
e.g. by \citet{2011ApJ...741...56D,2011ApJ...741...57D}, we have changed the computational setup, which
is listed briefly in Table~\ref{tab:local}. Despite the usage of cylindrical coordinates the setup is
in fact identical to a model used by \citet{2011ApJ...741...56D}. The very small thickness $H$ of the
disk and the small planet mass minimize curvature effects and make the problem more local.
The nonlinearity parameter for this local model is ${\cal M}  = 0.32$, which is similar to the standard case.
This quasi-local model has been run in a 2D and 3D setup using {\tt FARGO3D}. The 3D case has been run again in spherical polar
coordinates with the same spatial resolution as in the 2D setup of Table~\ref{tab:local}. For the
gravitational smoothing a length of two grid-cells has been chosen, which is equivalent here to $\epsilon = 0.06\,H$.
For the 2D simulations we use {\tt RH2D} and {\tt FARGO}, while 
for the 3D simulations we use {\tt NIRVANA} and {\tt FARGO3D}.  All these codes are based on the standard ZEUS-method
and are enhanced with the FARGO-speedup, see Appendix \ref{seca:codes} for details.
\begin{table}
	\caption{
		\label{tab:local}
		The setup for the alternative quasi-local model. 
 The parameters have been chosen according to \citet{2011ApJ...741...56D}.
	}
	\centering
	\renewcommand\arraystretch{1.2}
\begin{tabular}{lll}
	Parameter &  Symbol  &   Value  \\
	\hline
	mass ratio    &   $q = M_{\rm p}/M_*$  &  $3.2 \times 10^{-8}$   \\
	aspect ratio  &    $h= H/r$  &   $0.01$  \\
	nonlinearity parameter   &    ${\cal M}=q^{1/3}/h$  &   $0.32$    \\
	\hline
	potential smoothing  &   $\epsilon_\mathrm{p}$  &  $0.06 H$ \\
	radial range  &     $r_\mathrm{min}$ -- $r_\mathrm{max}$  &  $0.94$ -- $1.06$  \\
	angular range  &     $\phi_\mathrm{min}$ -- $\phi_\mathrm{max}$  &  $-0.32$ -- $0.32$ rad \\
	number of grid-cells    &  $N_r \times N_\phi  $      &  $384 \times 2048$  \\
	spatial resolution     &  $\Delta r$          &  $H/32$  \\
\hline
\end{tabular}
\end{table}

\begin{figure}
        \centering
        \includegraphics[width=\columnwidth]{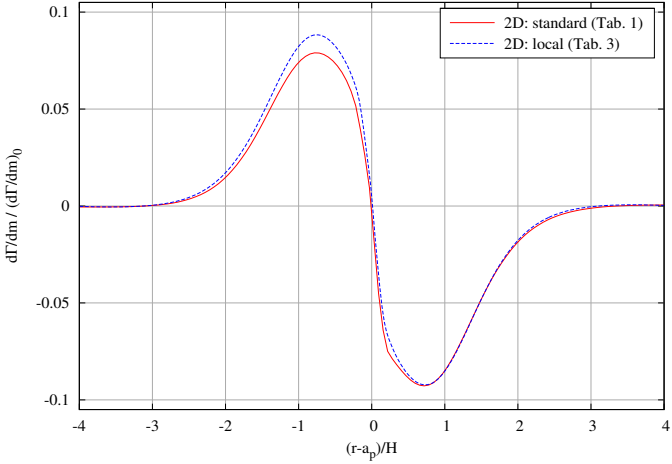}
        \caption{
                \label{fig:gam0-alt2D}
                The radial torque density in units of $(d\Gamma/d m)_0$.
                Compared is the standard setup with $q=6 \times 10^{-6}, h=0.05$ at 200 $T_\mathrm{orb}$ to
              the quasi-local model with $q=3.2 \times 10^{-8}, h=0.01$ at 30 $T_\mathrm{orb}$.
              The local calculation utilizes the {\tt FARGO3D}-code in the 2D setup. 
        }
\end{figure}

\begin{figure}
        \centering
        \includegraphics[width=\columnwidth]{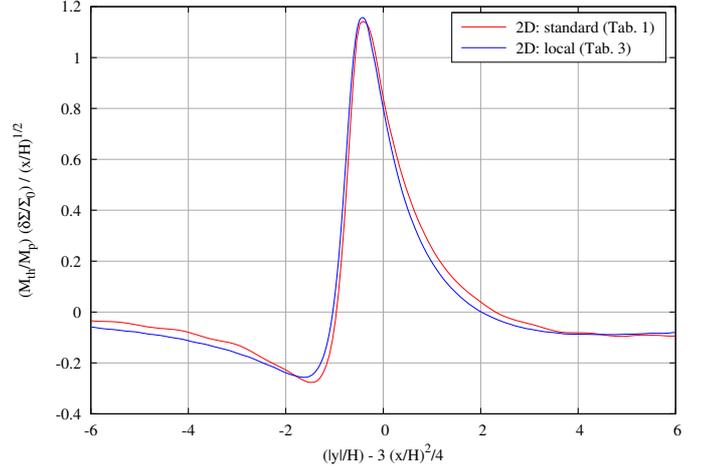}
        \caption{
                \label{fig:wakec-alt2D}
                Normalized azimuthal density profile of the outer wake at the radius $r_\mathrm{p} + 4/3 H$
               at 30 $T_\mathrm{orb}$. Compared is the standard setup with $q=6 \times 10^{-6}, h=0.05$ to
              the quasi-local model with $q=3.2 \times 10^{-8}, h=0.01$. 
        }
\end{figure}

\begin{figure}
        \centering
        \includegraphics[width=\columnwidth]{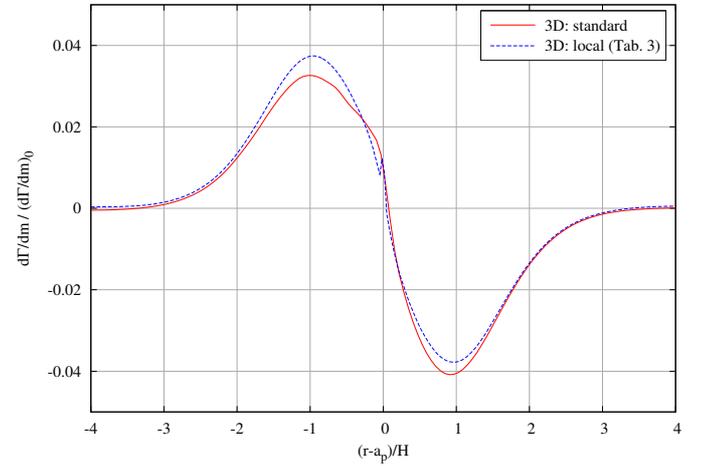}
        \caption{
                \label{fig:gam0-alt3D}
                The radial torque density in units of $(d\Gamma/d m)_0$ for full 3D models.
                Compared is the standard setup with $q=6 \times 10^{-6}, h=0.05$ at 30 $T_\mathrm{orb}$ (using
             {\tt NIRVANA}) to
              the  quasi-local model with $q=3.2 \times 10^{-8}, h=0.01$ at 15 $T_\mathrm{orb}$ (using {\tt FARGO3D}). 
        }
\end{figure}

\begin{figure}
        \centering
        \includegraphics[width=\columnwidth]{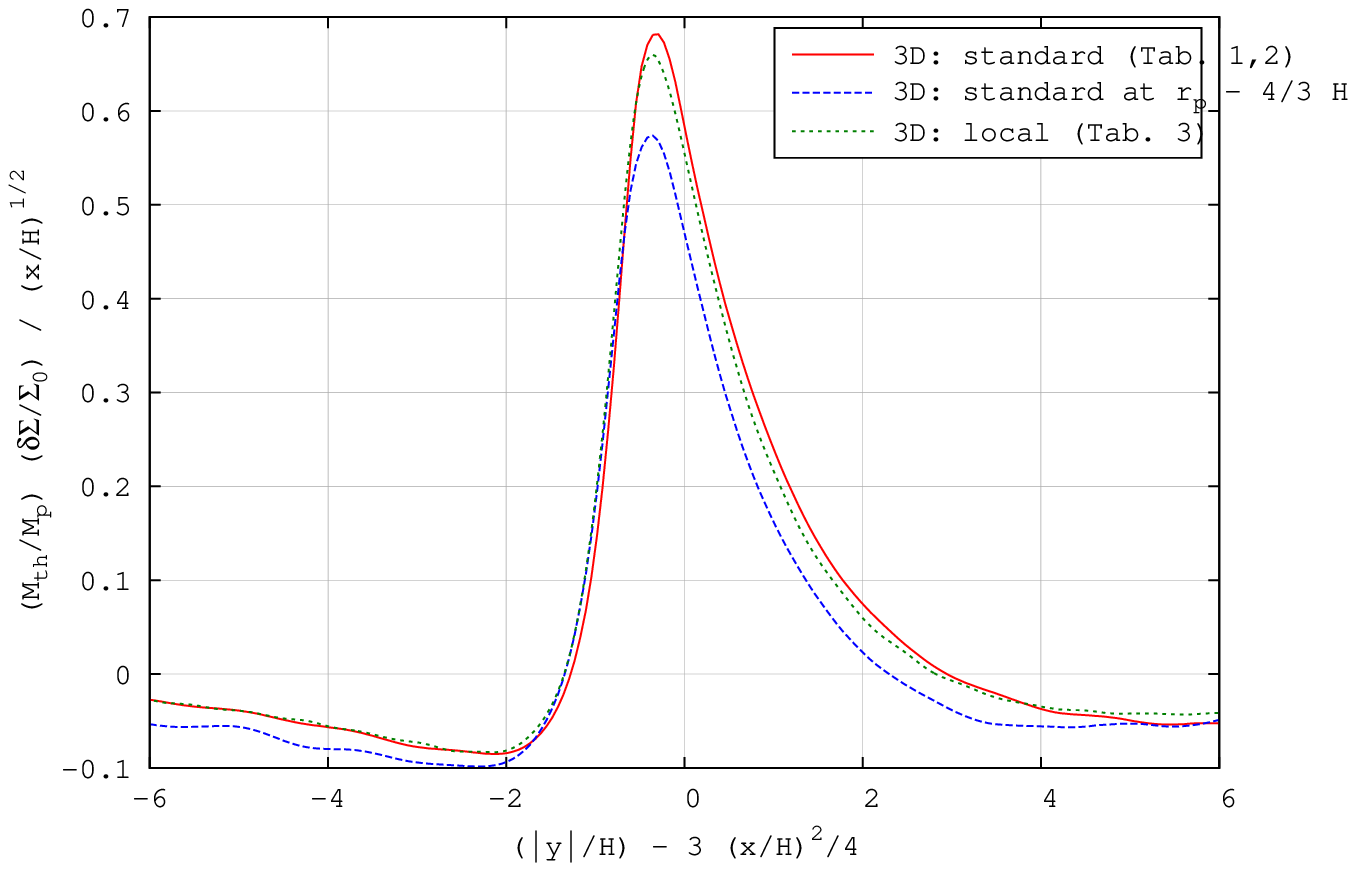}
        \caption{
                \label{fig:wakec-alt3D}
                Normalized azimuthal density profile of the outer and inner wake at the radii $r_\mathrm{p} \pm 4/3 H$
               for full 3D models. Compared is the standard setup with $q=6 \times 10^{-6}, h=0.05$ 
              at 30 $T_\mathrm{orb}$ for the outer and inner wake, to
              the quasi-local model with $q=3.2 \times 10^{-8}, h=0.01$ at 15 $T_\mathrm{orb}$, only at the outer wake.
        }
\end{figure}

In Fig.~\ref{fig:gam0-alt2D} we compare the torque density of the 2D standard model to the quasi-local model.
In a local setup any corotation torques saturate very quickly, possibly due to the very small (quasi-periodic) domain in the angular direction.
To match this condition, the standard model is shown here at $200 T_\mathrm{orb}$ when the corotation torques have
nearly saturated. 
The overall shape and magnitude of the two models is qualitatively in very good agreement,
which supports the scaling with $(d\Gamma/d m)_0$. 
For the local models a symmetric shape with respect to the location of the planet is expected, while 
for standard model the outer torques are larger in magnitude. This can explain some differences.

In Fig.~\ref{fig:wakec-alt2D} we compare the wake form of the standard model (as shown in Figs.~\ref{fig:wake0-std},\ref{fig:wakec-3D})
to the more local alternative model for the 2D setup. 
The two curves agree very well indeed, despite the huge difference in parameters for the planet mass and the disk scale height. 
We attribute the small differences to curvature effects.
We note that, due the local character of this setup, the curves for inner and outer wakes at $r_\mathrm{p} \pm 4/3$ 
look identical for the quasi-local model. 

In Fig.~\ref{fig:gam0-alt3D} we compare the torque density of the 3D standard model to the 3D quasi-local model.
As in the 2D case, now the overall shape and magnitude of the two models is again qualitatively in good agreement.
The local model shows a symmetric shape with respect to the location of the planet, as expected.
For both cases a similar reduction of amplitude in comparison to the 2D case is seen.

In Fig.~\ref{fig:wakec-alt3D} we compare the wake form of the standard model to the local alternative
model for the full 3D setup. This time the two curves for the outer wake agree again very well, despite the huge difference in parameters.
The profile for the inner wake of the standard model deviates from the outer wake as for the previous 2D setup.
For the local model, inner and outer wake are again identical, as expected.

\subsection{Adiabatic simulations}
\label{subsec:adi}

\begin{figure}
        \centering
        \includegraphics[width=\columnwidth]{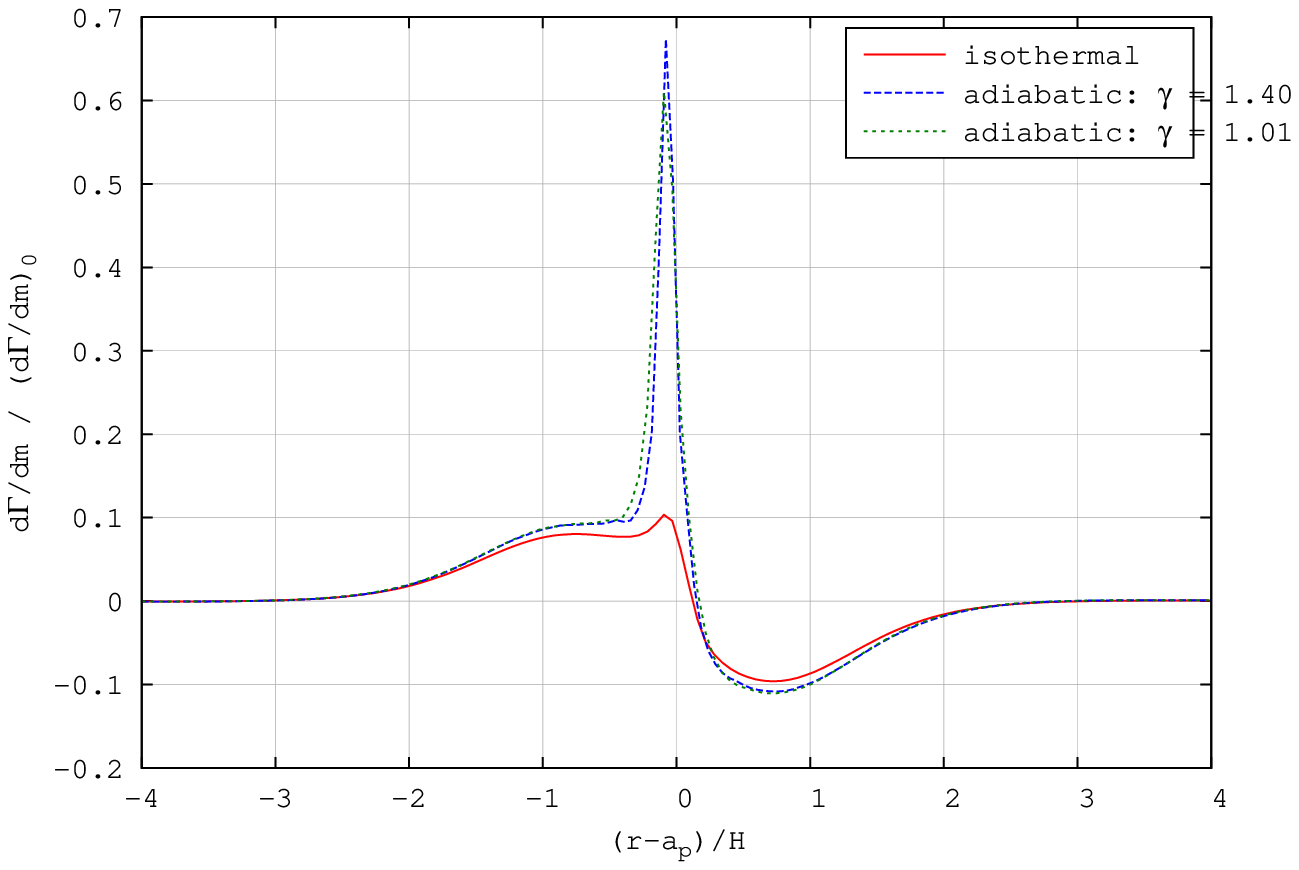}
        \caption{
                \label{fig:gam0-adi}
                The radial torque density in units of $(d\Gamma/d m)_0$ (see Eq.~\ref{eq:gamm0})
               at 30 $T_\mathrm{orb}$ for the 2D standard model for an isothermal and adiabatic setup using
               $\gamma = 1.01$ and $ \gamma = 1.40$.
                The units for $(d\Gamma/d m)_0$ and $H$ have been changed for the adiabatic runs, such
                 that $H \rightarrow \gamma H$.
        }
\end{figure}

\begin{figure}
        \centering
        \includegraphics[width=\columnwidth]{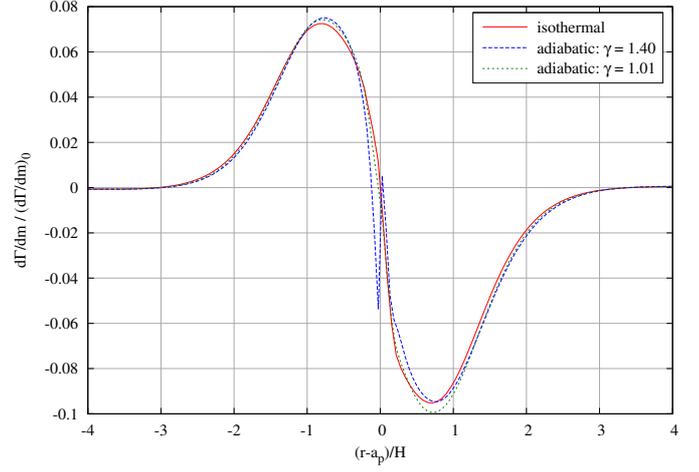}
        \caption{
                \label{fig:gam1-adi-500}
                The radial torque density in units of $(d\Gamma/d m)_0$ (see Eq.~\ref{eq:gamm0})
               at 500 $T_\mathrm{orb}$ for the 2D standard model for an isothermal and adiabatic setup using
               $\gamma = 1.01$ and $ \gamma = 1.40$.
                The units for $(d\Gamma/d m)_0$ and $H$ have been rescaled as in Fig.~\ref{fig:gam0-adi}.
        }
\end{figure}

The assumption of isothermality is only satisfied approximately in protoplanetary disks.
Because cooling times can be long, it maybe more appropriate to take into account the energy equation. 
To study the influence
of the equation of state on the outcome, we performed purely adiabatic simulations, that solve the energy
equation (Eq.~\ref{eq:energy}) together with an ideal equation of state. The result of such an approach is presented
in Fig.~\ref{fig:gam0-adi}, where the radial torque density is displayed for the standard isothermal model
together with two adiabatic models using $\gamma = 1.4$ and $1.01$, respectively.
The adiabatic results require rescaled units because the adiabatic sound speed is by a factor $\sqrt{\gamma}$ larger
than the isothermal one. Hence, the pressure scale length is increased by the same factor,
which enters (through $H$) the units for $(d\Gamma/d m)_0$ and $\Gamma_0$.
Obviously, there is a huge difference in the horseshoe torque between isothermal and adiabatic runs,
while the Lindblad contributions are similar, once correctly scaled. The adiabatic runs yield similar
results for the two $\gamma$ values throughout.
The strong torque enhancement in this adiabatic simulations comes from the entropy-related part of the
corotation torque which is driven by a radial gradient of entropy across the horseshoe region \citep{2008ApJ...672.1054B}.

\begin{figure}
        \centering
        \includegraphics[width=\columnwidth]{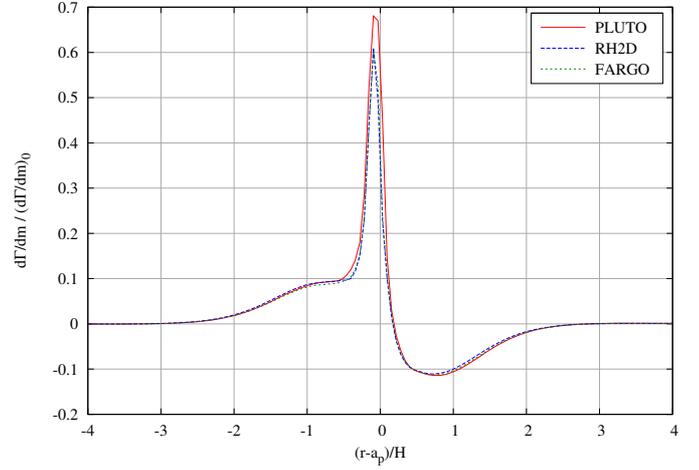}
        \caption{
                \label{figa:gamc-adi-codes}
                The radial torque density in units of $(d\Gamma/d m)_0$ (see Eq.~\ref{eq:gamm0})
               at 30 $T_\mathrm{orb}$ for the adiabatic standard model using an ideal equation of state
               with $\gamma = 1.01$. Three different codes have been used, {\tt RH2D} and {\tt FARGO} are
               2nd order upwind schemes and {\tt PLUTO} is a Riemann solver.
        }
\end{figure}

This result is interesting because sometimes an isothermal situation is mimicked with an adiabatic simulation
using a  $\gamma$-value very close to unity. In particular, this may be required by Riemann solvers that do not allow
to treat isothermal conditions. Our results show, that such an approach has to be treated very carefully,
as shown already by \citet{2008A&A...478..245P}. They argued that compressional heating near the planet plays an important
role in determining the torques.
Another reason lies in the fact that in an adiabatic situation the entropy is conserved along streamlines which
is not the case for isothermal flows. Reducing the value of $\gamma$ even further yields the same results. 
In general, an adiabatic flow with $\gamma \rightarrow 1$ approaches truly isothermal flow only in the
the case of a {\it globally} constant temperature.

After a few libration times the horseshoe region is well mixed, and the entropy and potential vorticity
gradients across the horseshoe regions are wiped out. Hence, the horseshoe torques disappear and the Lindblad
contributions remain. This situation is displayed in Fig.\ref{fig:gam1-adi-500} for an evolutionary time of
$500$ orbits. Now, the isothermal model agrees well with the adiabatic one.

We have applied several codes on the adiabatic setup as well.
In Fig.~\ref{figa:gamc-adi-codes} we display the same results for the adiabatic situation using $\gamma = 1.01$.
Again, all codes agree very well, even though now the numerical methodology is vastly different as some use a
second order upwind scheme ({\tt RH2D} and {\tt FARGO}) while {\tt PLUTO} uses a Riemann-solver.
Only very near to the planet the results differ slightly.

\begin{figure}
        \centering
        \includegraphics[width=\columnwidth]{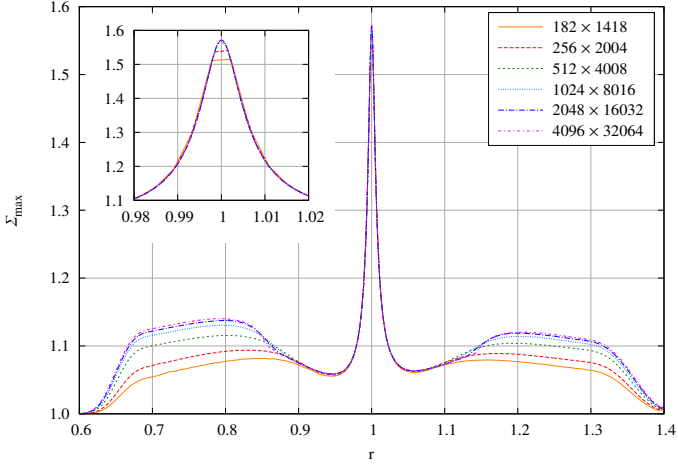}
        \caption{
                \label{fig:dens-max}
                The maximum of the density in the spiral wake as a function of radius for the 2D isothermal standard model
               at 30 $T_\mathrm{orb}$. Different numerical resolutions are shown using {\tt FARGO}.
        }
\end{figure}

\begin{figure}
        \centering
        \includegraphics[width=\columnwidth]{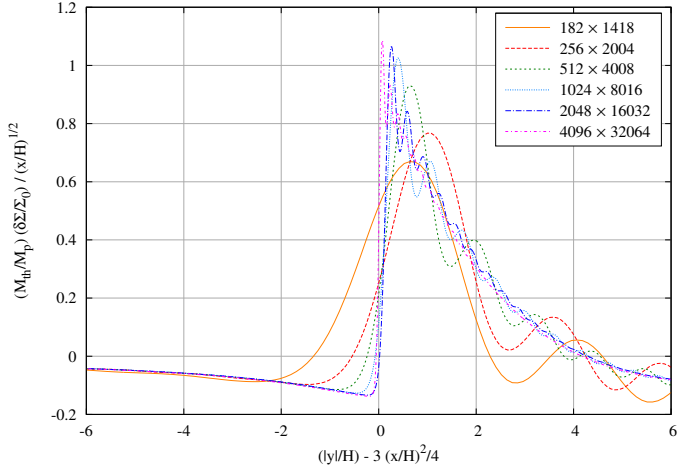}
        \caption{
                \label{fig:wakes5}
                Normalized azimuthal density profile of the outer wake at the radii $r_\mathrm{p} + 5 H$
               at 30 $T_\mathrm{orb}$ for the 2D isothermal standard model. Different numerical resolutions are shown.
        }
\end{figure}

\section{Shock formation}
\label{sec:shock}
For the damping of the wake it is important where the transition to a shock occurs. 
As a shock indicates a discontinuous change in the fluid variables, numerical codes often have difficulty to
resolve the structure in detail.
To analyze this, we plot in Fig.~\ref{fig:dens-max} the maximum density in the wake as a function of radius
for various resolutions of the computational grid. At the radius of the planet the density obviously has its maximum and it drops on both sides.
The previous curves for the wake profile have been taken near the minimum value of the density maximum.
Here, all resolutions show an identical maximum of the wake amplitude. Hence, as we have demonstrated in Sect.~\ref{subsec:std-resolution}, the
form of the wake does not depend very strongly on resolution at a distance of $|x| = 4/3 H$.

Further away from the planet, beyond a distance $|x| \gtrsim 2 H$ the
curves begin to differ for the various resolutions. This is clearly an indication for non-convergence of the simulations.
We attribute this to the formation of a shock wave. 
Indeed, at a distance $x_\mathrm{s} \approx 2 H$ from the location of the planet
the speed of the wake becomes supersonic with respect to the local Keplerian flow.
The criterion as  given by \citet{2001ApJ...552..793G} indicates for our nonlinearity parameter,
${\cal M}=0.36$, a shock formation at a distance of $~\approx 2.9 H$ from the planet, consistent with our findings.
At very high spatial resolutions, i.e. above a grid resolution of 64 grid-cells per
scale height (1024$\times$8016), the curves begin to converge in the shock region.
Far away from the planet, beyond $|x| \approx 6 H$ ($r=1.3$) the damping action of the boundary condition begins to set in,
and the curves coincide again.

In Fig.~\ref{fig:wakes5} the azimuthal density profile is shown at a radial location $r= r_\mathrm{p} + 5 H$ = 1.25.
At this location the wake is expected to have turned into a shock wave.
Due to the trailing nature of the wake we define the variable $y$ here slightly different as before through
\[
    y = (\phi - \phi_\mathrm{p}) \, r \,.
\]
From the figure it is obvious that the wake has turned into a shock at this location.
At our standard resolution (256$\times$2004) there is no indication for a shock front.
The overall form is very smooth with the presence of large oscillations behind the wake.
With increasing numerical resolution the shock becomes better and better resolved, but only at the very highest
resolution the wake turns into a discontinuous jump. The oscillations behind the front diminish and move closer to the front with
increasing resolution. Numerical experiments show that these oscillations can be damped out by increasing the
strength of viscosity. However, this smears out the shock front as well. It has been suggested that these oscillations are
due to the used numerical scheme and occur for very weak shocks \citep{2010arXiv1012.0266R}.

\section{Summary}
\label{sec:summary}
Through a series of 2D and 3D simulations using different computational methods and codes we have
explored in detail the numerical requirements for studies of the planet-disk problem.
In our analysis we focus on the torque density acting on the planet and the structure of the wake generated
by the planet.

With respect to the applicability of the fast orbital advection algorithm, FARGO, we have shown that
it leads to consistent numerical results that agree extremely well with non-FARGO studies.
The achievable gain in speed can be significant. For the setup used here we found a speed-up of more
than a factor of 10. The method works well in the presence of embedded planets, does not show
any signs of unsteady behavior, and can be applied in two or three spatial dimensions.
As it is applicable in conjunction with magnetic fields as well, new possibilities with respect
to numerical studies of turbulent accretion disks open up \citep{Pluto2012}.

Concerning the treatment of the gravitational potential of embedded planets, we extend previous studies 
\citep{2002A&A...387..605M,2012A&A...541A.123M} to very low mass planets in extremely thin disks.
We confirm that, for physical reasons, in 2D simulations the planetary potential has to be smoothed
with about $\epsilon = 0.6\,H $ -- $0.7\,H$. Models where the gravitational force is obtained directly through a
vertical integration yield always reasonable agreement with full 3D simulations.
The usage of very small smoothing lengths below $\epsilon = 0.6\,H$ in 2D simulations is not recommended,
because then the forces in the vicinity of the planet are strongly overestimated, which results in an unphysical
enhancement of the torque and too strong wakes.

Through a careful resolution study, we show that the smooth wake structure at distances smaller than about
$2\,H$ of the planet can be resolved well, and consistently, already with very low resolution of 8 to 16 cells per scale height.
The results are clearly converged for 32 grid-cells per $H$. For larger distances from the planet, the spiral wake turns into a shock wave
and much higher resolution may be required. We found, that around a resolution of about 100 grid-cells per $H$ convergence can be achieved.
Because this high resolution is only required near the spiral shocks, and the flow is relatively smooth outside,
numerical methods that adaptively refine this crucial region may be the method of choice in the future. 

For adiabatic flows we confirm earlier findings \citep{2008A&A...478..245P}
that the unsaturated horseshoe drag shows a strong deviation from the isothermal case.
Using the appropriate scaling the adiabatic corotation torques are independent of $\gamma$ and do not converge to
the isothermal case, even in the limit $\gamma \rightarrow 1$. Hence, the procedure of modelling the isothermal case with
simulations of $\gamma$ close to unity, has to be treated with care.
In the final saturated case, where all the corotation effects have been wiped out, isothermal and adiabatic
results agree perfectly, once the correction to the sound speed has been applied.

In Appendix \ref{subsec:std-resolution} we have shown that we do not find an additional timestep criterion
due to the planetary potential and we also have not noticed any unstable evolution in the case of using the
orbital advection. The question why in the simulations using the {\tt ATHENA}-code instabilities occur \citep{2011ApJ...741...56D}
may be connected to the treatment of orbital advection in that code \citep{2010ApJS..189..142S} which is apparently different from
the implementation in the {\tt FARGO}-code. One should also notice that in such simulations the conservative treatment
of Coriolis forces is mandatory to properly conserve angular momentum \citep{1998A&A...338L..37K}.   

We have demonstrated that the planet-disk interaction problem may be regarded as a very good test to validate an implementation
of orbital advection, because it admits a nearly analytic solution to which a code output can be compared. This is not the case for
simulations of turbulent disks, where no such known solutions exist.
We hope, that the presented results and comparison simulations may serve as a useful reference for other researcher in this field.
\begin{acknowledgements}
 Tobias M\"uller received financial support from the Carl-Zeiss-Stiftung.
 Wilhelm Kley acknowledges the support of the German Research Foundation (DFG) through grant KL 650/8-2                
 within the Collaborative Research Group FOR 759: {\it The formation of Planets: The Critical First Growth Phase}.
 Some simulations were performed
        on the bwGRiD cluster in T\"ubingen, which is funded by the Ministry for Education and Research of Germany and
        the Ministry for Science, Research and Arts of the state Baden-W\"urttemberg, and the cluster of the
        Forschergruppe FOR 759 "The Formation of Planets: The Critical First Growth Phase" funded by
        the Deutsche Forschungsgemeinschaft.
   Pablo  Ben\'\i tez-Llambay acknowledges the financial support of CONICET and the computational resources provided by IATE.
  We acknowledge fruitful discussions with Ruobing Dong and Roman Rafikov.
\end{acknowledgements}

\appendix

\section{The codes}
\label{seca:codes}

For our comparison simulations we utilized the following codes:

\noindent{\tt NIRVANA}: In its original (FORTRAN) version a ZEUS-like second order upwind scheme \citep{1997CoPhC.101...54Z}, with
the option of fixed nested grids and magneto-hydrodynamics (MHD). It can be used in two or three dimensions and can use
different coordinate systems. Recently, it has been improved to include radiative transport and the FARGO-treatment \citep{2009A&A...506..971K}.

\noindent{\tt RH2D}: A two-dimensional radiation hydrodynamics code for different coordinate systems, originally developed for treating the
boundary layer in accretion disks \citep{1989A&A...208...98K}, and later adapted to the planet disk problem
\citep{1999MNRAS.303..696K}.

\noindent{\tt FARGO}: A two-dimensional, special purpose code for disk simulations that first featured the FARGO-algorithm
\citep{2000A&AS..141..165M}. The code is publicly available at: {\tt http://fargo.in2p3.fr/},
and has been used frequently in planet-disk and related simulations.

\noindent{\tt FARGO3D}: A code based on similar algorithms as the standard {\tt FARGO}-code, but aimed at being more versatile, 
as it includes Cartesian, cylindrical and spherical geometries, in one, two or three dimensions, with arbitrary grid limits.
Its hydrodynamical core has been written from scratch, and it includes an MHD solver based on the method
of characteristics and constrained transport. It is parallelized using the Message Passing Interface (MPI) 
and a slab domain decomposition. It is intended in a nearby future to run distinctly on clusters
of CPUs or GPUs, and it will be made publicly available as the successor of the {\tt FARGO}-code.

\noindent{\tt PLUTO}: A multi-dimensional Riemann-solver based code for MHD
flows \citep{2007ApJS..170..228M}, which can be used in the purely hydrodynamic setup as well.
Additionally, it has been empowered recently by the FARGO-algorithm \citep{Pluto2012}.
{\tt PLUTO} is also freely available at: {\tt http://plutocode.ph.unito.it}.
  
The first 3 codes in the list have been used and described in an earlier code comparison project
on the planet-disk problem \citep{2006MNRAS.370..529D}. There, more massive planets of Neptune and
Jupiter mass embedded in viscous and inviscid disks have been studied for a large number of codes, and
the focus was on the gap structure of the disk and the total torques have been analyzed.  

\section{Timestep limitation in the presence of gravity}
\label{sec:time-gravity}
Numerically, we expect that possibly gravity might cause problems if, due to the gravitational acceleration $g$,
a parcel of material travels more than about half a gridcell of length $\Delta x$ in one timestep $\Delta t$.
This requires the additional gravitational criterion
\begin{equation}
\label{eq:deltag}
     \Delta t_\mathrm{G}  \leq \left(\frac{\Delta x}{g}\right)^{1/2} \,.
\end{equation}
Using now the smoothed planetary potential of Eq.~(\ref{eq:planet_2dpot}) we find that the maximum force is given
by
\begin{equation}
     g_\mathrm{max}  = \frac{G M_\mathrm{p}}{\epsilon^2} \, k
\end{equation}
with $k = 2/3^{3/2} \approx 0.4$. 
To obtain the strongest limitation on $\Delta t$ we substitute $g_\mathrm{max}$ in Eq.~(\ref{eq:deltag}) and obtain
\begin{equation}
\label{eq:deltagx}
     \Delta t_\mathrm{G}  \leq \Omega_\mathrm{K}^{-1} \, {\cal M}^{-1/2} \, 
       \left(\frac{2 \Delta x \, \epsilon^2}{k \, H^3} \right)^{1/2} \,.
\end{equation}
We compare this limit now to the regular Courant condition when using orbital advection
which is given by
\begin{equation}
     \Delta t_\mathrm{C}  =  \frac{\Delta x}{c_\mathrm{s}} \,,
\end{equation}
and find
\begin{equation}
     \frac{\Delta t_\mathrm{G}}{\Delta t_\mathrm{C}}  =  {\cal M}^{-1/2} 
       \left(\frac{2 \epsilon^2}{k H \Delta x} \right)^{1/2} \,.
\end{equation}
If there should be no additional timestep limitation generated by the gravity then this ratio should be
larger than one. Writing now for the grid resolution $\Delta x = H / N$ 
we obtain finally that
\begin{equation}
\label{eq:limitN}
       N  \geq  \frac{k}{2} \, \frac{H^2}{\epsilon ^2} \, {\cal M}
\end{equation}
for stability.
With $k=0.4$, $\epsilon = 0.1 H$, and ${\cal M} = 0.36$ we find for the necessary resolution
$N \approx 10$. This is indeed fulfilled even for our lowest resolution.
We point out that this limit formally only applies to flows without pressure (dust). If around the planet the envelope
is hydrostatic, no additional criterion is required. Switching on the planetary potential slowly will ensure stability
throughout the evolution as will an initial atmosphere around the planet \citep{2012ApJ...755....7D}. 

\section{The 3D hydrodynamic equations in a rotating frame}
\label{seca:eqn-rotframe}
For reference we state here the 3D hydrodynamic equations in a rotating coordinate frame.
In a coordinate system rotating with the (constant) angular velocity ${\bf \Omega}$,
omitting pressure terms, any external forces (eg. gravitation) and viscosity,
the momentum equation reads:
\begin{equation}
  \doverd{{\bf u}}{t} + {\bf u} \nabla {\bf u} =
     - 2 {\bf \Omega} \times {\bf u}
      + \frac{1}{2} \nabla \left[({\bf \Omega} \times {\bf r})^2\right]
\end{equation}
We now use spherical polar coordinates ($r, \varphi, \theta$),
where $r$ is the radial coordinate, $\varphi$ is the azimuthal angle,
and $\theta$ is the usual polar coordinate measured from the $z-$axis.
NOTE, that we use in this Appendix the same symbol $r$ for the spherical radial coordinate.
For rotation around the $z$-axis, ${\bf \Omega} = \Omega {\bf e}_z$, 
the individual equations are:    
\begin{equation}  \label{u_r}
 \doverd{u_r}{t} + {\bf u} \nabla u_r =
     \frac{1}{r} \left(u_{\varphi}^2 + u_\theta^2 \right)
        + r \Omega^2 \sin^2 \theta  + 2 u_{\varphi} \Omega \sin \theta
\end{equation}
\begin{equation}  \label{u_phi_sp}
 \doverd{u_{\varphi}}{t} + {\bf u} \nabla u_{\varphi} =
    - \frac{u_r u_{\varphi}}{r}
    - \frac{u_\theta u_{\varphi} \cot \theta}{r}
    - 2 \Omega \left( \sin \theta u_r  + \cos \theta u_\theta \right)
\end{equation}
\begin{equation}  \label{u_theta}
 \doverd{u_{\theta}}{t} + {\bf u} \nabla u_{\theta} =
    - \frac{u_r u_{\theta}}{r}
    + \frac{u_\varphi^2 \cot \theta}{r}
    + 2 \Omega u_\varphi \cos \theta
    + \Omega^2 r \sin \theta \cos \theta
\end{equation}
Introducing the angular velocity $\omega$ through
\begin{equation}
   u_\varphi = r \sin \theta \, \omega
\end{equation}
we may write for the three equations (\ref{u_r} - \ref{u_theta})
\begin{equation}
 \doverd{u_r}{t} + {\bf u} \nabla u_r
    =
     \frac{u_\theta^2}{r}
     + r \sin^2 \theta \left( \omega  + \Omega \right)^2
\end{equation}
\begin{equation} \label{u_phi1_sp}
 \doverd{u_{\varphi}}{t} + {\bf u} \nabla u_{\varphi} =
    - \left( \omega + 2 \Omega \right)
      \left( \sin \theta u_r  + \cos \theta u_\theta \right)
\end{equation}
\begin{equation}  \label{u_theta1}
 \doverd{u_{\theta}}{t} + {\bf u} \nabla u_{\theta} =
    - \frac{u_r u_{\theta}}{r}
     + r \sin \theta \cos \theta \left( \omega  + \Omega \right)^2
\end{equation}
One sees that in the radial and meridional ($\theta$) momentum
equation only the centrifugal part ($\omega + \Omega$) and in the
angular momentum ($\varphi$) equation only the Coriolis term
($2 \Omega$) occurs.

\subsection{Conservative treatment of Coriolis Terms in Angular Momentum
equation}
Defining the {\it total} specific angular momentum
\begin{equation}
    h_t  = r^2 \sin^2 \theta \left( \omega + \Omega \right)
\end{equation}
and using the continuity equation (in 3D) we may write for
the angular momentum equation (\ref{u_phi1_sp})
\begin{equation} \label{h_t}
 \doverd{\rho h_t}{t} + \nabla \cdot \left( \rho h_t {\bf u} \right) = 0
\end{equation}
Expanding $h_t$ this may be written:
\begin{equation} \label{eq:angmom}
 \doverd{\left[ \rho r \sin \theta
  \left( u_\varphi + r \Omega \sin \theta \right) \right]}{t}
    +
 \nabla \cdot \left[ \rho r \sin \theta
    \left( u_\varphi + r \Omega \sin \theta \right)
    {\bf u}  \right]  = 0
\end{equation}
The validity of this last equation (\ref{eq:angmom})
can be easily checked by expanding the terms and make use of the
continuity equation. Then one arrives at the equation (\ref{u_phi1_sp}).
In a numerical method that evolves $u_\varphi$ the equation
(\ref{eq:angmom}) should be used to solve the angular momentum transport conservatively.

\section{Comparing to linear results}
\begin{figure}
        \centering
        \includegraphics[width=\columnwidth]{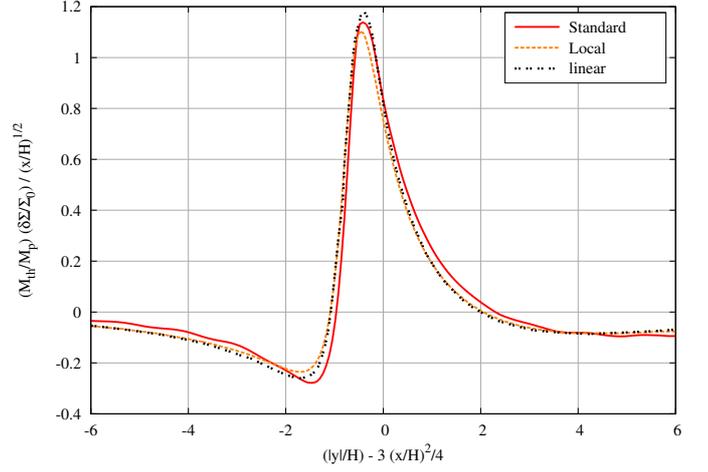}
        \caption{
                \label{fig:wakec-std-dong}
                Normalized azimuthal density profile of the outer wake at the radius $r_\mathrm{p} + 4/3 H$
               at 30 $T_\mathrm{orb}$. Compared is the standard setup with $q=6 \times 10^{-6}, h=0.05$ and
              the quasi-local model with $q=3.2 \times 10^{-8}, h=0.01$ to the linear theoretical results
             of \citet{2001ApJ...552..793G}.
        }
\end{figure}

After submission of the original manuscript, Ruobing Dong generously supplied us with the data
of the linear results of \citet{2001ApJ...552..793G}. In Fig.~\ref{fig:wakec-std-dong} we compare their
data to our results for the 2D simulations using the standard setup of Tab.~\ref{tab:standard} and
the quasi-local setup of Tab.~\ref{tab:local}. The overall agreement of our full nonlinear results with
the linear case is very good. The small differences between the results are comparable to what \citet{2011ApJ...741...56D}
found in their study. Please note, that their vertical scaling differs by a factor of 3/2.

\bibliographystyle{aa}
\bibliography{wake}
\end{document}